\documentclass[twocolumn]{aastex62}

\usepackage{lipsum}
\usepackage{mathtools}

\newcommand{\teff}{T_{\mathrm{eff}}}
\newcommand{\age}{{\mathrm{age}}}

\newcommand{\logg}{\log g}
\newcommand{\feh}{[\mathrm{Fe}/\mathrm{H}]}
\newcommand{\alphafe}{[\alpha/\mathrm{Fe}]}

\DeclarePairedDelimiter\abs{\lvert}{\rvert}%
\DeclarePairedDelimiter\norm{\lVert}{\rVert}%

\makeatletter
\let\oldabs\abs
\def\abs{\@ifstar{\oldabs}{\oldabs*}}
\let\oldnorm\norm
\def\norm{\@ifstar{\oldnorm}{\oldnorm*}}
\makeatother

\graphicspath{{./}{figures/}}

\received{February 13, 2023}
\revised{May 5, 2023}
\accepted{May 8, 2023}
\submitjournal{Astrophysical Journal}

\shorttitle{HALO7D III: Chemical Properties}
\shortauthors{McKinnon et al.}

\begin{document}

\title{HALO7D III: Chemical Abundances of Milky Way Halo Stars from Medium Resolution Spectra}

\correspondingauthor{Kevin McKinnon}
\email{kevin.mckinnon@ucsc.edu}

\author[0000-0001-7494-5910]{Kevin A. McKinnon}
\affiliation{Department of Astronomy \& Astrophysics, University of California, Santa Cruz, 1156 High Street, Santa Cruz, CA 95064, USA}

\author[0000-0002-6993-0826]{Emily C. Cunningham}
\altaffiliation{NASA Hubble Fellow}
\affiliation{Department of Astronomy, Columbia University, 550 West 120th Street, New York, NY, 10027, USA}
\affiliation{Center for Computational Astrophysics, Flatiron Institute, 162 5th Ave, New York, NY 10010, USA}

\author[0000-0002-6667-7028]{Constance M. Rockosi}
\affiliation{Department of Astronomy \& Astrophysics, University of California, Santa Cruz, 1156 High Street, Santa Cruz, CA 95064, USA}

\author[0000-0001-8867-4234]{Puragra Guhathakurta}
\affiliation{Department of Astronomy \& Astrophysics, University of California, Santa Cruz, 1156 High Street, Santa Cruz, CA 95064, USA}

\author[0000-0002-9933-9551]{Ivanna Escala}
\altaffiliation{Carnegie-Princeton Fellow}
\affiliation{Department of Astrophysical Sciences, Princeton University, 4 Ivy Lane, Princeton, NJ 08544}
\affiliation{The Observatories of the Carnegie Institution for Science, 813 Santa Barbara St, Pasadena, CA 91101}

\author[0000-0001-6196-5162]{Evan N. Kirby}
\affiliation{Department of Physics and Astronomy, University of Notre Dame, Notre Dame, IN 46556, USA}

\author[0000-0001-6146-2645]{Alis J. Deason}
\affiliation{Institute for Computational Cosmology, Department of Physics, University of Durham, South Road, Durham DH1 3LE, UK}

\begin{abstract}
The Halo Assembly in Lambda Cold Dark Matter: Observations in 7 Dimensions (HALO7D) survey measures the kinematics and chemical properties of stars in the Milky Way (MW) stellar halo to learn about the formation of our Galaxy. HALO7D consists of Keck II/DEIMOS spectroscopy and Hubble Space Telescope-measured proper motions of MW halo main sequence turn-off (MSTO) stars in the four CANDELS fields. HALO7D consists of deep pencil beams, making it complementary to other contemporary wide-field surveys. We present the $\feh$ and $\alphafe$ abundances for 113 HALO7D stars in the Galactocentric radial range of $\sim10-40$~kpc along four separate pointings. Using the full 7D chemodynamical data (3D positions, 3D velocities, and abundances) of HALO7D, we measure the velocity anisotropy, $\beta$, of the halo velocity ellipsoid for each field and for different metallicity-binned subsamples. We find that two of the four fields have stars on very radial orbits, while the remaining two have stars on more isotropic orbits. Separating the stars into high, mid, and low $\feh$ bins at $-2.2$~dex and $-1.1$~dex for each field separately, we find differences in the anisotropies between the fields and between the bins; some fields appear dominated by radial orbits in all bins while other fields show variation between the $\feh$ bins. These chemodynamical differences are evidence that the HALO7D fields have different fractional contributions from the progenitors that built up the MW stellar halo. Our results highlight the additional information available on smaller spatial scales compared to results from a spherical average of the stellar halo. 
\end{abstract}

\keywords{Milky Way stellar halo (1060), Milky Way Galaxy (1054), Stellar abundances (1577), Astrostatistics (1882)}

\section{Introduction} \label{sec:intro}

Owing to their long dynamical timescales, galactic stellar halos are long-lived structures that preserve information about their origins. Within a $\Lambda$CDM cosmology, galaxies are built up by merger events, each of which can contribute stars to the halo. The positions, kinematics, and chemical properties of halo stars thus reveal a galaxy's mass assembly history and information about the dwarf galaxy progenitors that contributed to it. By studying the stellar halo of our home Galaxy, we seek to place the Milky Way (MW) in its cosmological context.

The chemodynamical properties of halo stars are powerful for constraining masses, star formation rates and efficiencies, and accretion times of progenitors as well as the total mass and shape of the MW gravitational potential \citep[e.g.,][]{Eggen_1962,Searle_1978,Bullock_2005,Helmi_2008}. While the positional clustering of stars from an accreted satellite is eventually washed out, the kinematic coherence of accreted debris persists for much longer periods of time. Stellar atmospheric chemical abundances are relatively stable over a star's lifetime, and they are a direct result of the environment in which it was formed; interstellar medium enrichment levels, star formation rates, and formation lifetimes of a galaxy all impact the chemical makeup of star-forming gas as a function of time.

As $\alpha$ elements (i.e. O, Ne, Mg, Si, S, Ar, Ca, Ti) can be produced at early times in core-collapse events at relatively constant rates with iron, high $\alphafe$ stars tend to be formed at early times in a galaxy's star-forming life. After enough time \citep[e.g. $\sim 330$~Myr as measured by][]{Maoz_2010}, and assuming sufficient star formation, Type Ia supernovae ``turn on'' and create much of the iron we see in the universe. This causes the $\alphafe$ ratio to drop as $\feh$ increases \citep{Wallerstein_1962,Tinsley_1980}. The mass-metallicity relation of local dwarf galaxies \citep{Kirby_2013,Kirby_2017,Kirby_2020} reveals that the metallicity distribution function (MDF) of a galaxy is tied to its mass; this arises because more massive systems have deeper potential wells that are able to retain a greater fraction of their enriched gas from supernovae. While this relationship was determined for local dwarf galaxies that are observed today, recent work \citep[e.g.][]{Leethochawalit_2018,Leethochawalit_2019,Naidu_2022} has explored how these mass-metallicity relations were different at earlier cosmic times\@. The star formation and quenching time of a progenitor system are impacted during its accretion, but the chemical properties of the accreted stars are linked to the mass, star formation rate, and formation lifetime of their birth environment; this inter-relatedness enables the inferences of progenitor properties from stellar halo chemical abundances \citep[e.g.,][]{Lee_2015,Hasselquist_2021,Horta_2022,Cunningham_2022}.

At present time, a stellar halo is comprised of stars from many different progenitor systems. 
Simulations of purely accreted stellar halos \citep[e.g.][]{Bullock_2005,Robertson_2005,Font_2006a,Font_2006b,Font_2006c} have average stellar halo abundances driven by their merger histories \citep[e.g.][]{Johnston_2008}, with average $\alphafe$ tracking accretion time of infalling dwarf galaxies and $\feh$ tracking the mass/luminosity of those dwarfs. Recent work with state-of-the-art simulations \citep[e.g.,][using the \textit{Latte} suite of FIRE-2 simulations; \citet{Hopkins_2015,Hopkins_2018,Wetzel_2016}]{Horta_2022} have further refined our understanding of how the distributions of present-day chemodynamical observables of stellar halo stars are dictated by the accretion times and masses of disrupted dwarfs. 

In addition to stars accreted from dwarf galaxy mergers, in-situ stars -- those formed in the potential well of the host galaxy -- can be heated onto orbits similar to accreted halo stars during merger events. As a result, the halo population is a combination of in-situ stars and those accreted during the many mergers a galaxy experiences throughout its history. \citet{Zolotov_2009}, for example, finds evidence from simulations that the in-situ halo fraction of the inner regions traces how quiescent the recent merger history has been; more recent mergers tend to cause the inner halo in-situ fraction to decrease\footnote{While their definition of ``in-situ stars'' is different than ours, the dependence on fractional contribution of in-situ halo stars as a function of merger history should be similar.}. Using data from the H3 survey \citep{Conroy_2019a}, \citet{Naidu_2020} find that the in-situ halo fraction drops from 60\% to 5\% of the total halo mass when going from Galactocentric radii of 8~kpc to 20~kpc.

For this paper, ``in-situ halo'' refers to the progenitor high-$\alpha$ disk that was kinematically heated through early merger events \citep{Nissen_2010,Bonaca_2017,Bonaca_2020,Haywood_2018,DiMatteo_2019,Amarante_2020}. This is the so-called ``Splash'' of \citet{Belokurov_2020}. It is thus chemically similar to the thick disk -- that is, peaked towards relatively metal-rich $\feh$ around $-0.5$~dex \citep{Naidu_2020} -- but consists of stars on more isotropic orbits instead of predominantly circular ones. \citet{Belokurov_2020} show that their ``Splash'' sample has less net prograde rotation and larger scatters in all velocity components compared to their ``Thick Disk'' sample.

Thanks in large part to the \textit{Gaia} survey \citep{Gaia_collaboration_2018},  our current picture of the MW's formation history is becoming clearer: aside from the recent interactions with Sgr and the LMC/SMC, the MW halo seems to have had a relatively quiescent recent history. Recent work has shown that the inner $\sim25$~kpc of the stellar halo is dominated by a single massive progenitor called Gaia-Sausage-Enceladus (GSE) \citep{Belokurov_2018,Helmi_2018,Haywood_2018}. The GSE is relatively metal-rich, with a peak $\feh$ of $\sim-1.2$~dex \citep{Naidu_2020}, and radially biased orbits. The GSE has a net rotation, $\langle v_\phi \rangle$, that is consistent with 0$~{\rm km \ s}^{-1}$ \citep{Belokurov_2020}, and is estimated to have come from a 4:1 mass ratio merger approximately 10 Gyr ago \citep{Helmi_2018}. While the GSE and in-situ halo dominate the bulk of the inner regions of the stellar halo, there have been many other substructures identified over the last five years \citep[e.g.][]{Myeong_2019,Koppelman_2019,Yuan_2020,Naidu_2020,Belokurov_2020}. There is growing evidence that 92\% to 99\% of the MW stellar halo stars can be associated with one of the currently known progenitors \citep{Naidu_2020}. 

These detailed inventories of the MW stellar halo have helped constrain answers to questions about our Galaxy, such as the approximate formation history, fractional contributions from different progenitors, and range of progenitor properties. Many of these studies rely on \textit{Gaia}-based parallaxes and proper motions. Because of \textit{Gaia}'s apparent magnitude limit of $G\sim 20$~mag, this means that these samples are either focused on nearby main sequence (MS) halo stars when studying the local halo or more distant evolved stars when studying the distant halo. These evolved stars are intrinsically less numerous/spatially dense than their MS counterparts, which means that much of the distant-halo work has focused on average properties over large areas of the sky. 

The Halo Assembly in Lambda Cold Dark Matter: Observations in 7 Dimensions \citep[HALO7D;][]{Cunningham_2016,Cunningham_2019a,Cunningham_2019b} is complementary to contemporary MW stellar halo surveys because it targets 3D positions, 3D velocities, and abundances of main sequence turn-off stars at moderate halo radii ($10 < r < 40$~kpc). The HALO7D sample consists of Keck II/DEIMOS spectroscopy and HST-measured proper motions for stars in the apparent magnitude range $19 < m_v < 24.5$~mag, making it a deep complement to \textit{Gaia}-based surveys. This paper, the third in the series, measures the 7-th and final dimension of HALO7D stars: chemical abundances. The high spatial density of MS halo stars allows us to compare average chemodynamical properties along different lines of sight (LOS)\@. \citet{Cunningham_2019b}, for example, measure the halo velocity anisotropy along the four HALO7D LOS and find variations between the different fields. 

In this paper, we describe the HALO7D data set in Section \ref{sec:data}. We create a Bayesian spectrophotometric pipeline to measure chemical abundances ($\feh$ and $\alphafe$) and stellar parameters ($\teff$, $\logg$, $\age$, and distance) for MSTO stars without known distances in Section \ref{sec:methods} and present the resulting abundance measurements for the HALO7D sample in Section \ref{sec:results}. In Section \ref{sec:discussion}, we combine our abundances with previously-measured LOS velocities and proper motions from HALO7D \citep{Cunningham_2019a,Cunningham_2019b} to measure the variation in average chemodynamical properties -- such as net halo  rotation, $\langle v_\phi\rangle$, and the velocity anisotropy parameter, $\beta$ -- using different subsamples of the HALO7D stars. Our findings are summarized in Section \ref{sec:conclusion}. The detailed tests on the outputs of our chemical abundance pipeline are described in Appendix \ref{sec:fake_stars}, and then validated against well-studied globular clusters in Appendix \ref{sec:gc_comparison}. We show the statistical significance of our kinematic measurements in Appendix \ref{sec:anisotropy_pipeline}.

\section{Data} \label{sec:data}
The HALO7D dataset consists of HST-measured proper motions and Keck II/ DEIMOS spectroscopy for 199 main sequence turn-off (MSTO) MW halo stars in four Cosmic Assembly Near-infrared Deep Extragalactic Legacy Survey \citep[CANDELS;][PIs: S. Faber, H. Ferguson]{Grogin_2011,Koekemoer_2011} fields: COSMOS, GOODSN, GOODSS, and EGS\@. These fields are located at high Galactic latitudes -- meaning they have minimal foreground contamination from the MW disk -- and they are not located in regions of previously-known streams or substructure (e.g. Sagittarius).
While the stars that make up the HALO7D sample lie within the same footprints as the CANDELS fields \citep[see Figure 1 of][]{Cunningham_2019a} -- and we refer to them using the same field names -- it should be noted that the HALO7D dataset does not include every MSTO star found in the CANDELS fields. The first paper in this series \citep{Cunningham_2019a} presents the DEIMOS spectroscopic dataset (see their Section 2) and measures line-of-sight velocities with a hierarchical Bayesian pipeline called \texttt{velociraptor}. The second paper \citep{Cunningham_2019b} presents proper motions measured from multi-epoch HST imaging (see their Section 2 and Table 2 for a description of the \textit{HST} Programs) and characterizes the halo velocity ellipsoid. 

A detailed description of the HALO7D sample and the selection process are presented in the first two papers in this series. To summarize the relevant information, the fields were chosen because of their deep, multi-epoch HST astrometry and photometry, which enables proper motion measurements to much fainter magnitudes than \textit{Gaia}. HALO7D's velocity sample consists of stars in the $19 < m_{v} < 24.5$~mag range and are generally blue to minimize the impact of contamination by foreground disk stars; to see the full CMD selection criteria, please see Section 2.2 and Figures 2 and 3 of \citet{Cunningham_2019a}. HALO7D is thus complementary to previous studies because it is able to measure kinematics and chemical abundances for individual main sequence stars at halo distances with high enough spatial sampling density to measure halo properties along individual lines-of-sight instead of measuring sky-averaged properties versus Galactocentric radius.

\begin{figure}[h]
\begin{center}
\includegraphics[width=\linewidth]{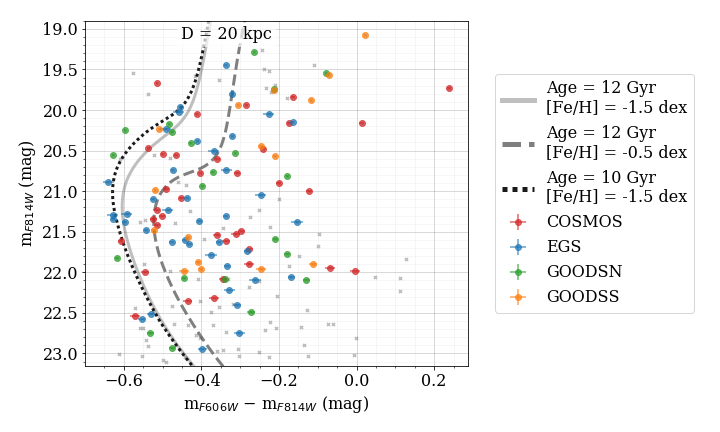}
\caption{Color-magnitude diagram of the 113 HALO7D stars with measured LOS velocities and chemical abundances in HST filters (STMAG)\@. Three MIST isochrones at a distance of 20~kpc with typical MW halo properties are shown in black/grey to guide the eye in this region of color-magnitude space; an increase (decrease) in distance causes the isochrones to move down (up) vertically to fainter (brighter) apparent magnitudes. The faint grey dots show the 86 HALO7D stars that do not have measured abundances.}
\label{fig:HALO7D_CMDs}
\end{center}
\end{figure}

\begin{figure}[h]
\begin{center}
\includegraphics[width=\linewidth]{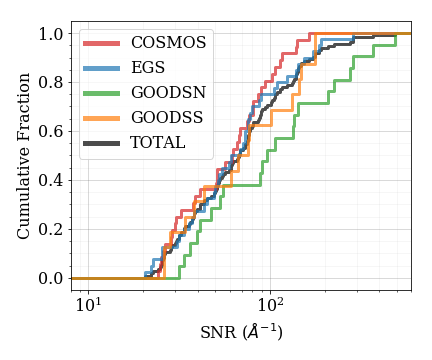}
\caption{Cumulative histograms of the combined spectral signal-to-noise ratios of stars in the chemistry sample of HALO7D (colored points in Figure \ref{fig:HALO7D_CMDs}). The black line represents the total sample. By design, GOODSN had more observations than the other fields and thus shows higher combined SNRs\@. The median SNRs for each field and the total sample are found in the text.}
\label{fig:HALO7D_SNRs}
\end{center}
\end{figure}

The Keck II/DEIMOS spectroscopic observations were collected between March 2014 and April 2017 using the 600 line/mm grating with a $7200~\rm \AA$ central wavelength configuration and 1'' slitwidth. These medium resolution spectra ($R\sim 2000$) consist of 8192 pixels, pixel spacing of $\sim 0.65~\rm \AA$/pixel, and covering a typical wavelength range of $\sim 5000-10000~\rm \AA$. Each target was typically observed for $\sim 5-6$ hours over the course of this time period, with a minimum of 2 and a maximum of 33 individual visits per target. The raw spectroscopic data were reduced with the \texttt{spec2d} pipeline produced by DEEP2 at UC Berkeley \citep{Cooper_2012}. Table \ref{t:observations} shows a summary of the observations in each field, Figure \ref{fig:HALO7D_CMDs} shows a color-magnitude diagram for the 199 stars in HALO7D with the colored points denoting the 113 stars for which we are able to measure abundances. Figure \ref{fig:HALO7D_SNRs} shows cumulative histograms of the combined spectral signal-to-noise ratio (SNR) for the stars with abundances in each field, where we define the combined SNR using a quadrature sum of the SNRs of the individual observations: $$\mathrm{SNR}_{\mathrm{combined}} = \left( \sum_{i}^{n_{obs}} \mathrm{SNR}_i^2 \right)^{1/2}.$$ The median combined spectral SNRs are 62.2, 64.8, 95.8, and 71.3$~\rm \AA^{-1}$ for COSMOS, EGS, GOODSN, and GOODSS respectively and 67.2$~\rm \AA^{-1}$ for the complete sample. 

\begin{table*}[t]
\centering
\caption{Summary of Targets in CANDELS Fields with line-of-sight velocities and proper motions from \citet{Cunningham_2019a} and \citet{Cunningham_2019b}.
\label{t:observations}}
\begin{tabular}{cccccccc}
\hline\hline
            &  $l$   & $b$       & Area\footnote{The listed field area is the area covered with HST multi-epoch imaging.}             & N Halo Stars & Catalog References\\
Field       & (deg) & (deg)     & (arcmin$^{2}$)   & with $v_{3D}$ \\ \hline 
\multicolumn{1}{c}{COSMOS}   &  236.8 & 42.1  & 288         & 81  & \citet{Nayyeri_2017,Muzzin_2013}\\
\multicolumn{1}{c}{GOODSN}  &  125.9 & 54.8  & 166         & 32   & \citet{Barro_2019}\\
\multicolumn{1}{c}{GOODSS}  &  223.6 & $-54.4$ & 160         & 20  & \citet{Guo_2013} \\
\multicolumn{1}{c}{EGS}      &  96.4  & 60.4  & 384         & 66  & \citet{Stefanon_2017,Barro_2011}\\ 
\multicolumn{1}{c}{Total}      &  --  & --  & 998           & 199 \\ 
\hline
\end{tabular}
\end{table*}

\section{Measuring Stellar Parameters and Abundances} \label{sec:methods}

We create a Bayesian pipeline that uses stellar photometry and spectroscopy to measure chemical abundances ($\feh$, $\alphafe$), stellar parameters ($\teff$, $\logg$, age), and distances for our HALO7D stars. This technique relies on the library of MIST isochrones\footnote{\url{https://waps.cfa.harvard.edu/MIST/index.html}} \citep{Dotter_2016,Choi_2016,Paxton_2011,Paxton_2013,Paxton_2015} and a set of synthetic model spectra (described by \citet{Escala_2019} in the blue, $4100-6300~\rm \AA$, and \citet{Kirby_2008,Kirby_2011_PASP} in the red, $6300-9100~\rm \AA$). The model spectra are generated with the MOOG spectral-synthesis software \citep{Sneden_1973}, using the ATLAS9 model atmospheres \citep{Kurucz_1993,Kirby_2011_PASP}. Each synthetic spectrum is defined by a set of ($\teff,~\logg,~\feh,~\alphafe$) values. For these spectra, $\alphafe$ is the total $\alpha$-element abundance of a stellar atmosphere, where O, Ne, Mg, Si, S, Ar, Ca, and Ti are the $\alpha$-elements we consider.

Many techniques that measure chemical abundances and stellar parameters from spectra using synthetic model spectra typically study populations of stars that have photometrically-constrained distances \citep[for example]{Kirby_2010,Escala_2019} or parallax measurements \citep{Conroy_2019a, Cargile_2020}. This means that distance-degenerate parameters, such as $\logg$, are able to be fixed and a smaller region of parameter space needs to be explored. For the HALO7D sample, however, we deal with MSTO stars with unknown distances. The absolute magnitudes of stars near the MSTO vary much more than their colors, which means that the stars have large distance uncertainties.  As a result, we must explore larger regions of parameter space and consider the relationships between distance and the possible stellar parameters. To help constrain our abundance fits, we use the MIST isochrones, measured properties of the MW stellar halo, and photometric observations of the stars in our sample to create multi-dimensional, multi-modal prior distributions on the parameters of interest for each star. 

\subsection{Generating Prior Distributions of Stellar Parameters}  \label{ssec:generating_priors}

In building our priors, we use MIST isochrones in the range of $0.1 <\mathrm{age} < 14.6~\mathrm{Gyr}$ and $-4.0 <\feh < 0$~dex in steps of 0.2~Gyr and 0.02~dex respectively. The current version of the publicly available MIST isochrones are only for $\alphafe = 0$~dex, so we are not able to include this parameter in the isochrone-derived prior distributions; we assume a flat prior in the range of $-0.8 < \alphafe < +1.2$~dex. These isochrones are then assigned weights based on average properties of the MW stellar halo, such as an assumed metallicity distribution function (MDF) and age distribution. Our particular choices of parameters for the halo priors are shown in Table \ref{t:prior_distributions}. The $\feh$ distribution is chosen to have a moderately metal-rich peak and a width that is not overly constraining that is consistent with recent halo studies \citep[e.g.,][]{Belokurov_2018,Helmi_2018,Mackereth_2019,Conroy_2019b}. Similarly, the age distribution comes previous studies such as \citet{Kalirai_2012} and \citet{Bonaca_2020}. As we show in Appendix \ref{sec:gc_comparison}, the abundances we measure are fairly insensitive to our choice of priors in the spectral SNR range of our data. 

\begin{table*}[t]
\centering
\caption{MW halo prior distributions for abundance pipeline.}
\begin{tabular}{cc}
\hline \hline
Distribution                                   & Functional Form \\ \hline
\multicolumn{1}{c|}{$p(\feh)$}            & $\mathcal{N}(-1.5~\mathrm{dex},1.0~\mathrm{dex})$          \\
\multicolumn{1}{c|}{$p(\mathrm{age})$}            & $\mathcal{N}(12~\mathrm{Gyr},2.0~\mathrm{Gyr})$          \\
\multicolumn{1}{c|}{$p(\alphafe)$}            & $\propto 1$, for $-0.8 < \alphafe < 1.2$          \\
\multicolumn{1}{c|}{$p(\mathrm{mass} | \feh, \mathrm{age})$}            & \citet{Kroupa_2001} IMF,   $\propto k \left(\frac{\mathrm{mass}}{M_\odot}\right)^{-\alpha}$ with
        $\begin{cases}
            k=25,~\alpha = 0.3, & \mathrm{mass} < 0.08 M_\odot \\
            k=2,~\alpha = 1.3, & \mathrm{mass} < 0.5 M_\odot \\
            k=1,~\alpha = 2.3, & \mathrm{mass} > 0.5 M_\odot
        \end{cases}$ \\
\multicolumn{1}{c|}{$p(\mu)$}  & $\propto D^3\left(\frac{R_q}{27 \mathrm{kpc}}\right)^{-\alpha}$ with
        $\begin{cases}
            \alpha = 2.3, & R_q < 27\, \mathrm{kpc} \\
            \alpha =4.6, & R_q \geq 27\, \mathrm{kpc} 
        \end{cases}$ \\ \hline
\end{tabular}
\label{t:prior_distributions}
\end{table*}

The points within a particular isochrone each have an initial mass, $M_{F606W}-M_{F814W}$ color, $M_{F814W}$ absolute magnitude, $\teff$, and $\logg$\@. The isochrone points are weighted by integrating a \citet{Kroupa_2001} initial mass function (IMF) over the range of masses within an isochrone to account for uneven mass spacing. 

Until now, the isochrones have been weighted only by properties that are generic to a halo population, but we now create unique prior distributions for the stellar parameters of each star. We apply additional weights to the isochrone points for each star separately using the observed photometry. Each star has observed apparent HST magnitudes, $m_{F606W}$ and $m_{814W}$, and corresponding uncertainties, $\sigma_{F606W}$ and $\sigma_{F814W}$, which we have de-reddened using the dust maps of \citet{Schlegel_1998} through the \texttt{dustmaps} package in \texttt{Python} \citep{Green_2018}. 

To give the isochrone points weights based on distances/absolute magnitudes, we use the MW stellar halo density profile of \citet{Deason_2011}:
\begin{equation*} \label{eq:gal_dist_weight}
\begin{split}
        \frac{dN}{dV} \propto
        \begin{cases}
            \left( \frac{R_q}{27\, \mathrm{kpc}}\right)^{-2.3}, & R_q < 27\, \mathrm{kpc} \\
            \left( \frac{R_q}{27\, \mathrm{kpc}}\right)^{-4.6}, & R_q \geq 27\, \mathrm{kpc}
        \end{cases}
\end{split}
\end{equation*}
\noindent where $R_q^2 = x^2+y^2+(z/q)^2$ with $q=0.59$. Accounting for volume elements and the Jacobian between distance and distance modulus, the distance modulus prior is thus $p(\mu)\propto D^3\cdot \frac{dN}{dV}$\@. We marginalize over the distance modulus by taking equally spaced values in $\mu$ that correspond to distances between 0.1~kpc and 500~kpc and comparing the observed stellar magnitudes to the implied apparent magnitude of the isochrone point at a given distance modulus. Overall, the weight of a given isochrone point ends up as: 
\begin{equation*} 
\begin{split}
p(\mathrm{p}_i | \feh, \mathrm{age}&) \propto p(\mathrm{mass}_i | \feh,\mathrm{age}) \cdot \sum_{j=0}^{n_{\mu}} [p(\mu_j) \cdot\\
&\mathcal{N}(M_{i,F606W} +\mu_j | m_{F606W},\sigma_{F606W})\cdot\\
&\mathcal{N}(M_{i,F814W}+\mu_j | m_{F814W},\sigma_{F814W})]
\end{split}
\end{equation*}
\noindent where $\mathrm{p}_i$ is the $i$-th point in an isochrone defined by $(\feh, \mathrm{age})$ which has absolute magnitudes \newline $(M_{i,F606W},M_{i,F814W})$\@. 

When comparing the MIST isochrones to HST photometry of stars in a handful of nearby globular clusters, we noticed that many of the isochrones required color offsets as large as $\sim 0.02$~mag to have better agreement with the data.
To allow for a potential mismatch between the MIST isochrones and the halo stars, we also marginalize over color offsets between $-0.02$~mag and $0.02$~mag. This is much larger than the typical photometric uncertainty of stars in the HALO7D sample (median uncertainty in $m_{F606W}-m_{F814W}$ of $0.006$~mag), so this marginalization has a relatively large effect in increasing the width of the prior distributions of stellar parameters.

For each star, the total weighting for a given isochrone point is thus a product of its isochrone $\feh$ and $\age$ weighting, its mass weighting, and the weighting from marginalizing over the distance modulus while comparing the photometry.

At a given color in an isochrone, there are generally three possible distances corresponding to a star being on the main sequence, the subgiant branch, or the horizontal branch. This results in our prior distributions having three local maxima: one peak for each of the possible phases in a star's evolution. Because it is generally more efficient to sample posterior distributions that are singly-peaked, we break each isochrone up into these three phases. The prior probability of a particular phase in any given isochrone is the fraction of the total weight of the isochrone in that phase. In this work, $\mathrm{phase} = 0$ corresponds to MS stars, $\mathrm{phase} = 1$ corresponds to the sub-giant branch, and $\mathrm{phase} = 2$ corresponds to the horizontal branch\footnote{In the MIST parlance, our $\mathrm{phase} = 0$ is also their $\mathrm{phase} = 0$, our $\mathrm{phase} = 1$ is their $\mathrm{phase} = 2$, and our $\mathrm{phase} = 2$ is their $\mathrm{phase} = 3$ and $\mathrm{phase} = 4$\@.}.

Finally, for each phase of each isochrone, we generate a 3D prior distribution on ($\teff$, $\logg$, and $M_{F814W}$) by passing the weighted isochrone points to a kernel-density estimator (KDE) with a Gaussian kernel. We compute the KDE width by measuring a standard deviation and mean of each of the parameters ($\teff$, $\logg$, and $M_{F814W}$) in the set of weighted isochrones, and normalize the measurements so that each individual parameter's distribution corresponds to a unit Gaussian. The Gaussian width used in the kernel density estimator is chosen to be 0.1, such that the width is 10\% of the standard deviation in each parameter. This KDE approach has the benefit of allowing combinations of $\teff$, $\logg$, and $M_{F814W}$ that do not fall perfectly on the isochrone, meaning we are less reliant on the isochrones perfectly capturing the relationships between stellar parameters. 

Overall, the prior probabilities of the stellar parameters are described by:
\begin{equation} \label{eq:stellar_param_prior}
\begin{split}
p(\vec \theta_*) =& p(\teff,\logg,\feh,\alphafe,M_{814W},\mathrm{age},\mathrm{phase})\\
\propto&p(\alphafe) \cdot p(\feh) \cdot p(\mathrm{age}) \cdot p(\mathrm{phase} | \feh, \mathrm{age}) \cdot\\
& p(\teff,\logg,M_{814W} | \feh, \mathrm{age}, \mathrm{phase})
\end{split}
\end{equation}
\noindent where $p(\teff,\logg,M_{814W} | \feh, \mathrm{age}, \mathrm{phase})$ is calculated by evaluating the KDE generated from the isochrone points with that particular $\feh$, age, and phase: $KDE_{\feh, \mathrm{age}, \mathrm{phase}}(\teff,\logg,M_{814W})$\@.

\subsection{Preprocessing of Spectra} \label{ssec:preprocessing_spectra}

Before the spectral observations can be used in our abundance pipeline, we must shift each of the spectra to the rest frame, continuum-normalize, identify useful wavelength regions, and characterize a few quality-of-observation parameters such as the line-spread function (LSF)\@. For our chemical abundance pipeline, we do not coadd the multiple spectral observations of a given star into a single observation. Instead, we model each observation simultaneously. We choose this approach because the spectral observations were taken over the course of years and can have vastly different observing conditions (e.g. seeing, line spread functions, wavelength solution offsets), which causes their coadded spectrum to have washed-out/hard-to-model features. This section is quite technical, so readers who are more interested in the big-picture steps of the abundance pipeline may choose to skip ahead to Section \ref{ssec:fitting_process} or to the results in Section \ref{sec:results}.

\subsubsection{Measuring Line Spread Functions} \label{sssec:measuring_LSFs}

As a star's light passes through the atmosphere to the telescope, its light is spread out by the seeing we measure during data collection. It is adequate for our purposes to assume the resulting shape is a Gaussian with a width that is the $\mathrm{FWHM_{seeing,i}}/2.355$, where $i\in\{1,\dots,n_{obs}\}$ refers to the spectral observation number. Accounting for the pixel scale and anamorphic factor allows us to convert this into a width in units of Angstroms in the spectral dimension. Because of the 1" width of the slits in the DEIMOS mask, any light outside of this width is truncated. The star's truncated-Gaussian light is further smoothed out as it passes through the instrument and lands on the CCD\@. To characterize this additional amount of instrument smoothing as a function of wavelength, we use arc lamp exposures\footnote{We use Kr, Xe, Ar, Ne for our red arcs, and the same elements plus Hg, Cd, Zn for our blue arcs.}. 

We identify peaks in the arc lamp spectrum and then fit those peaks with a top-hat function convolved by a Gaussian, where the Gaussian width is unknown and the top-hat width is set by the size of the slit. This is because the arc lamp light fully illuminates each slit, producing a top-hat shape, before it passes through the instrument. In particular, we are interested in measuring the smoothing Gaussian widths. The resulting widths as a function of wavelength are approximately quadratic with a minimum near the chip gap. For each individual observation, we measure this quadratic function on the blue- and red-chips separately, giving the additional smoothing of the instrument as a function of wavelength. The final LSF$_i(\lambda)$ for the $i$-th observation is then the seeing-defined truncated Gaussian convolved with the wavelength-dependent instrument smoothing. In cases where the seeing is very good, such that virtually all of the star's light is completely inside of the slit (e.g. $3\sigma_{\mathrm{seeing}} < \mathrm{slitwidth}/2$ which implies a seeing FWHM of $\sim0.39$" for our slitwidth of 1"), the resulting LSF$_i(\lambda)$  is a Gaussian with a total width that is the quadrature sum of the seeing Gaussian and the instrument smoothing. In most cases, however, the seeing is large enough relative to the slitwidth that we find it better to use the truncated-Gaussian-then-smoothed model instead. During the fitting process, we allow for the instrument smoothing function to change by a multiplicative factor between 0.5 and 1.5 to not be overly constraining. We place a prior on this multiplicative factor that is a Gaussian centered at 1 and has a width of 0.1 for each observation and allow the blue and red data to have different multiplicative factors:
$$\mathrm{blue~mult}_{\mathrm{phase},i} \sim \mathcal{N}(1.0,0.1)$$
$$\mathrm{red~mult}_{\mathrm{phase},i} \sim \mathcal{N}(1.0,0.1)$$

\subsubsection{Shift to Stellar Restframe and Pixel Masking} \label{sssec:shift_to_restframe}

Next, we shift each of the $n_{obs}$ observations of a single star to stellar restframe using the \texttt{velociraptor}-measured LOS velocities from \citet{Cunningham_2019a}. Specifically, we use the median LOS velocity from each spectral observation's posterior distribution. The data wavelengths are then restricted to $4100-9100~\rm \AA$ because that is the coverage of the synthetic model spectra we will compare to. We also mask out a few additional wavelength regions, some before shifting to restframe and others after, as listed in Table \ref{t:masked_regions}; these include telluric features, poorly-modeled (or un-modeled) features in the synthetic model spectra, and regions where the wavelength solution of the spectral reductions are unstable. 

For our spectroscopic setup with DEIMOS, there are not many arc lamp lines at particularly blue wavelengths, which makes anchoring the wavelength solution difficult in this region. We find that the wavelengths on the red-chip (i.e. $\lambda > 7200~\rm \AA$) are robust and that the blue-chip wavelengths are quite reliable down to $\lambda \sim 5500~\rm \AA$, but they are often unstable below $\lambda \sim 5000~\rm \AA$\@. Using a handful of stars with previously-measured stellar parameters and high SNR spectra, we cross-correlate with the best-fit synthetic model spectrum at different wavelength locations to re-measure the wavelength solution and compare it to the output of \texttt{spec2d}. Because the wavelength solution we measure is using the stellar features at all wavelengths, this process is able to anchor the solution using lines at the bluest wavelength where arc lines are not available. We find that the offset as a function of wavelength is approximately linear for $\lambda > 5000~\rm \AA$ on the blue-chip, with the size of the maximum offset being $\sim 3~\rm \AA$\@. 

With these lessons in mind, we decide to mask out wavelengths less than $5000~\rm \AA$ for all spectra because the wavelength offset function tends to become non-linear in this region. During the stellar parameter measuring process that we will discuss in Section \ref{ssec:fitting_process}, we fit each spectral observation with a linear wavelength offset function on the blue-chip wavelengths to correct for these issues. Ideally, we would use more terms in the wavelength offset function.  However, we must balance the number of parameters being fit per observation with computation time; we find that a linear correction function is a good compromise. For spectra with SNR less than $20~\rm \AA^{-1}$, we mask out wavelengths less than $5500~\rm \AA$ because this is where it becomes difficult to assess how well the wavelength offset has been measured. For the same reason, we drop individual spectral observations with SNR less than $3~\rm \AA^{-1}$\@. 

\begin{table}[h]\caption{Masked spectral regions.
\label{t:masked_regions}}
\centering
\begin{tabular}{cc}
\hline \hline
\multicolumn{2}{c}{\textbf{Masked Before Shift to Rest Frame}} \\
Name                                   & Region Masked (\AA) \\ \hline
\multicolumn{1}{c|}{Unstable Wavelength Solution\footnote{Very blue DEIMOS wavelengths tend to be extrapolations of a small number of arc lamp lines and are not easily fixed with our correction prescription described in Section \ref{sssec:wavelength_solution_correction}.}}            & $< 5000$          \\
\multicolumn{1}{c|}{B-Band}            & (6864, 7020)          \\
\multicolumn{1}{c|}{A-Band}            & (7570, 7713)          \\
\multicolumn{1}{c|}{Telluric Feature}  & (7157, 7325)          \\
\multicolumn{1}{c|}{Telluric Feature}  & (8123, 8356)          \\
\multicolumn{1}{c|}{Telluric Feature}  & (8235, 8275)          \\
\multicolumn{1}{c|}{Telluric Feature}  & $> 8933$             \\
\multicolumn{1}{c|}{Chip Gap}          & $\pm 20$ at gap              \\ \hline \hline
\multicolumn{2}{c}{\textbf{Masked After Shift to Rest Frame}}  \\
Name                                   & Region Masked (\AA) \\ \hline
\multicolumn{1}{c|}{MgH\footnote{This molecular absorption feature is masked for stars that have more than 5\% prior probability of $\teff < 5300$K because the synthetic models currently do not include this feature.}}      & (5115, 5125)          \\
\multicolumn{1}{c|}{Na D1, D2\footnote{This absorption feature is masked because of potential contamination from ISM absorption.}}      & (5885, 5910)          \\
\multicolumn{1}{c|}{Ca I ($\lambda$6343)}      & (6341, 6346)          \\
\multicolumn{1}{c|}{Ca I ($\lambda$6362)}      & (6356, 6365)          \\
\multicolumn{1}{c|}{H$\alpha$}         & (6559.797, 6565.797)          \\
\multicolumn{1}{c|}{K I ($\lambda$7665)}       & (7662, 7668)          \\
\multicolumn{1}{c|}{V I ($\lambda$8116, $\lambda$8119),}      & (8113, 8123)          \\
\multicolumn{1}{c|}{Poor Arcturus Model \footnote{From \citet{Kirby_2008}, this is a region where the model spectra were showed to poorly reproduce the spectral features of Arcturus.}}     & (8317, 8330)          \\
\multicolumn{1}{c|}{Ca II ($\lambda$8498)}     & (8488.023, 8508.023)  \\
\multicolumn{1}{c|}{Ca II ($\lambda$8542)}     & (8525.091, 8561.091)  \\
\multicolumn{1}{c|}{Ca II ($\lambda$8662)}     & (8645.141, 8679.141)  \\
\multicolumn{1}{c|}{Mg I ($\lambda$8807)}      & (8804.756, 8809.756)  \\ \hline
\end{tabular}
\end{table}

\subsubsection{Continuum Normalization} \label{sssec:continuum_norm}

We then measure an initial continuum estimate for each spectrum and for each possible stellar evolution phase, as defined in Section \ref{ssec:generating_priors}. We take the synthetic model defined by the median stellar parameters of the corresponding prior distribution, degrade it to the data quality using the LSF$_i(\lambda)$, and then divide the result from the observed spectrum. This gives an approximate continuum assuming a particular phase; we then smooth the resulting spectrum using a median boxcar of width 6.5$~\rm \AA$ (i.e. approximately 10 pixels) to limit the impact of outliers such as poorly-removed skylines. This median-binned spectrum is then fit with a cubic B-spline whose knots are spaced by 100$~\rm \AA$, yielding an estimate of a given observation's continuum assuming a particular phase, $c_{\mathrm{phase},i}(\lambda)$\@. This process is repeated for each spectral observation and for each phase.

\subsubsection{Identifying Useful Wavelength Regions and Corrections to Wavelength Solutions} \label{sssec:wavelength_solution_correction}

We also use these degraded model spectra to build prior distributions for the parameters of the linear wavelength offset function, $\Delta \lambda_{i}(\lambda) = m_{\lambda,i}\cdot \lambda+b_{\lambda,i}$\@. We cross-correlate the synthetic model with the spectral data in steps of $\sim ~100~\rm \AA$\@. This gives measures of the wavelength offset as a function of wavelength for a particular observation. We then fit a line to these offsets, taking the resulting fit's mean, $\vec \theta_{\lambda,i}$, and covariance matrix, $\pmb{V}_{\lambda,i}$, to define the prior distribution as a multivariate normal distribution, $$m_{\lambda,i}, b_{\lambda,i} \sim \mathcal{MVN}_2(\vec \theta_{\lambda,i} , 10\pmb{V}_{\lambda,i})$$ where we inflate the covariance matrix by a factor of 10 so as not to be overly constraining. 

Finally, we identify wavelength regions of the spectra that are particularly useful for measuring stellar parameters. Using the prior distributions to define a useful region of parameter space (e.g. 95\% prior probability), we are able to define a grid of equally spaced $\teff$, $\logg$, $\feh$, and $\alphafe$ values for each phase. For each set of parameters on the prior grid, we degrade the corresponding synthetic model to the data quality and compare the model spectra to each other as well as to the continuum-normalized spectral observations. This is done for two purposes. The first reason is to determine wavelength regions that do not change significantly across the parameter grid, which means they provide the least power for our likelihood measurements. We choose to mask the 10\% least useful pixels of each observation. This has the additional benefit of speeding up computations because we consider a smaller number of pixels; in total, the masking of poorly-modeled, telluric, and low-likelihood information regions leaves between $35 - 55$\% of the data pixels ($\sim3000 - 4500$ of the original 8129 pixels) for the likelihood measurements. The other reason is to better define the continuum for each observation given a particular phase. For each set of parameters in our prior grid, we divide the continuum-normalized data by the current synthetic model and then fit a B-spline with 100$~\rm \AA$-spaced knots to the result, which defines the continuum adjustment required to have the best agreement between that model and the data. The new continuum for each phase is taken to be the initial continuum multiplied by the new continuum adjustment of the model that had the minimum $\chi^2$ comparison. The parameters that define the best model (i.e. minimum $\chi^2$) for each phase is where we choose to begin our search of parameter space during the fitting process. 

\subsection{Fitting Spectra with Synthetic Models}  \label{ssec:fitting_process}
With our prior distributions and rest frame, continuum-normalized spectra in hand for each star, we begin the fitting process. The synthetic model spectra we compare to are defined by a set of $\teff,\, \logg,\, \feh,$ and $\alphafe$ values according to the model grid in Table \ref{t:model_grid}. In cases where we draw a set of parameters that don't lie directly on the model grid, we linearly interpolate from the nearest neighboring models using between 2 and $2^4$ nearest neighbors in ($\teff$,$\logg$,$\feh$,$\alphafe$) space. For each observation of a given star, the synthetic model is smoothed with the corresponding LSF$_i(\lambda)$ and re-sampled onto the data wavelengths. 

\begin{table}[h]
\caption{Model spectrum grid spacing by parameter value from \citet[Table 4]{Escala_2019}.
\label{t:model_grid}}
\centering
\begin{tabular}{cccc}
\hline \hline
Parameter                               & Min. Value                & Max. Value        & Step              \\ \hline
$\teff$ (K)          & 3500                         & 5600                 & 100                  \\
                                        & 5600                         & 8000                 & 200                  \\
$\logg$ (cm s$^{-2}$) & 0.0 ($\teff < 7000$ K)     & 5.0                  & 0.5                  \\
                                        & 0.5 ($\teff > 7000$~K)     & 5.0                  & 0.5                  \\
$\feh$                 & $-4.5$ ($\teff \leq 4100$~K) & $0.0$                  & 0.1                  \\
                                        & $-5$ ($\teff \leq 4100$ K)   & $0.0$                  & 0.1                  \\
$\alphafe$                              & $-0.8$                         & $+1.2$                  & 0.1                  \\ \hline
\end{tabular}
\end{table}

The posterior probability for our parameters of interest is:
\begin{equation} \label{eq:full_posterior}
    \begin{split}
        p(\vec \theta_*&,\pmb{\theta}_{spec} | \pmb{F}, \pmb{\Sigma}) \propto p(\vec \theta_*) \cdot p(\vec{\theta}_{spec}) \cdot p(\pmb{F} | \vec \theta_*,\pmb{\theta}_{spec}, \pmb{\Sigma}) \\
        \propto &p(\vec \theta_*) \cdot \prod_{i=1}^{n_{obs}}\left[ p(\vec{\theta}_{spec,i}) \cdot p(\vec f_i | \vec \theta_*,\vec{\theta}_{spec,i},\vec \sigma_i) \right]\\
        \propto &p(\vec \theta_*) \cdot \prod_{i=1}^{n_{obs}}\left[ p(\vec{\theta}_{spec,i}) \cdot \prod_{j=1}^{n_{pix}}p(f_{i,j} | \vec \theta_*,\vec{\theta}_{spec,i}, \sigma_{i,j}) \right]\\
    \end{split}
\end{equation}
\noindent where $i$ corresponds to the spectral observation number, $j$ corresponds to the pixel number within a spectrum, $\pmb{F} = (\vec f_1, \vec f_2, \dots, \vec f_{n_{obs}})$ are the fluxes of the measured spectra with corresponding uncertainties of $\pmb{\Sigma} = (\vec \sigma_1, \dots, \vec \sigma_{n_{obs}})$, $\pmb{\theta}_{spec} = (\vec \theta_{spec,1}, \dots, \vec \theta_{spec,n_{obs}})$ are the set of spectral parameters for all observations, and $\theta_*$ is the set of stellar parameters with $p(\theta_*)$ defined in Equation \ref{eq:stellar_param_prior}. Because the spectra for a given star are on a common wavelength array, $\lambda_j$ represents the wavelength of a given pixel, and $f_{i,j}$ is the flux measured in that pixel for spectrum $i$, with corresponding flux uncertainty $\sigma_{i,j}$\@.

The $\vec{\theta}_{spec,i}$ parameters include a blue- and red-side multiplier to the $LSF_i(\lambda)$, and the slope and intercept for the linear wavelength solution correction of that observation. Thus, the prior on the spectral parameters for each observation is:
\begin{equation*}
    \begin{split}
        p(\vec \theta_{spec,i}) = p(\mathrm{blue~mult}_{i})\cdot p(\mathrm{red~mult}_{i}) \cdot p(m_{\lambda,i},b_{\lambda,i})
    \end{split}
\end{equation*}
where $p(\mathrm{blue~mult}_{i})$ and $p(\mathrm{red~mult}_{i})$ are defined in Section \ref{sssec:measuring_LSFs} and $p(m_{\lambda,i},b_{\lambda,i})$ is defined in Section \ref{sssec:wavelength_solution_correction}.

The likelihood of a particular spectrum's flux measurement at a particular pixel, $p(f_{i,j} | \vec \theta_*,\vec{\theta}_{spec,i})$, comes from a comparison with the synthetic model that has had measured continuum applied. While the continuum we have measured in the previous section does a good job of normalizing each spectrum, we allow for one final continuum fit before evaluating the likelihood to have an optimal comparison. This is particularly important in cases where there is significant line blanketing at blue wavelengths because the original continuum definition may have removed the effects of line blanketing while trying to normalize the fluxes. To this end, for every draw of parameters, we smooth the corresponding synthetic model to the appropriate data quality, then divide it from the continuum-normalized data; we then coadd the remaining noise spectra from the different observations and fit the result with a final B-spline with 100$~\rm \AA$-spaced knots. This process captures any remaining large-scale variations that are required to have good agreement between a particular model and the data. 

In words, the parameter measurement proceeds as follows:
\begin{enumerate}
    \item Draw stellar parameters \newline ($\teff,~\logg,~\feh,~\alphafe,~M_{814W},~\mathrm{age},~\mathrm{phase}$) and spectral correction parameters ($\mathrm{blue~mult}_i$, $\mathrm{red~mult}_i$, $m_{\lambda,i}$, and $b_{\lambda,i}$ for each of the $i$ spectral observations) using Multivariate Normal proposal distributions;
    \item Read in the high-resolution synthetic model defined by the current stellar parameters;
    \item For each observation, smooth the synthetic model by LSF$_{i}(\lambda < \lambda_{\mathrm{chipgap}})\cdot \mathrm{blue~mult}_{\mathrm{phase},i}$ on the blue side and LSF$_{i}(\lambda > \lambda_{\mathrm{chipgap}})\cdot \mathrm{red~mult}_{\mathrm{phase},i}$ on the red side to get the data-quality model fluxes;
    \item Define the new wavelength vector for each observation using the current linear wavelength offset correction, $\Delta \lambda_{i}(\lambda) = m_{\lambda,i}\cdot \lambda + b_{\lambda,i}$, applied only to the blue side (i.e. $\Delta \lambda (\lambda > \lambda_{\mathrm{chipgap}}) = 0$);
    \item Re-sample the corresponding smoothed model onto this new wavelength array, giving $\vec m_i$;
    \item For each observation, divide the continuum-normalized observation, $\vec f_i/ \vec c_{\mathrm{phase},i}$, by the re-sampled smoothed model, $\vec m_i$, to get a noise spectrum that is centered at 1;
    \item Coadd the noise spectra together, and fit the result with a B-Spline to account for any missing continuum; this is the continuum adjustment vector $\vec a$;
    \item For each observation, the likelihood is then $p(\vec f_i | \vec \theta_*, \vec \theta_{spec,i}) = \prod_{j=1}^{n_{pix}} \mathcal{N}(f_{i,j} | a_{j}\cdot c_{\mathrm{phase},i,j}\cdot m_{i,j}, \sigma_{i,j})$;
    \item Measure the posterior probability for a current draw using Equation \ref{eq:full_posterior};
    \item Use Metropolis-Hastings criterion to accept or reject the drawn parameters;
    \item Repeat until the parameter samples have converged.
\end{enumerate}

\begin{figure*}[t]
\begin{center}
\includegraphics[width=\linewidth]{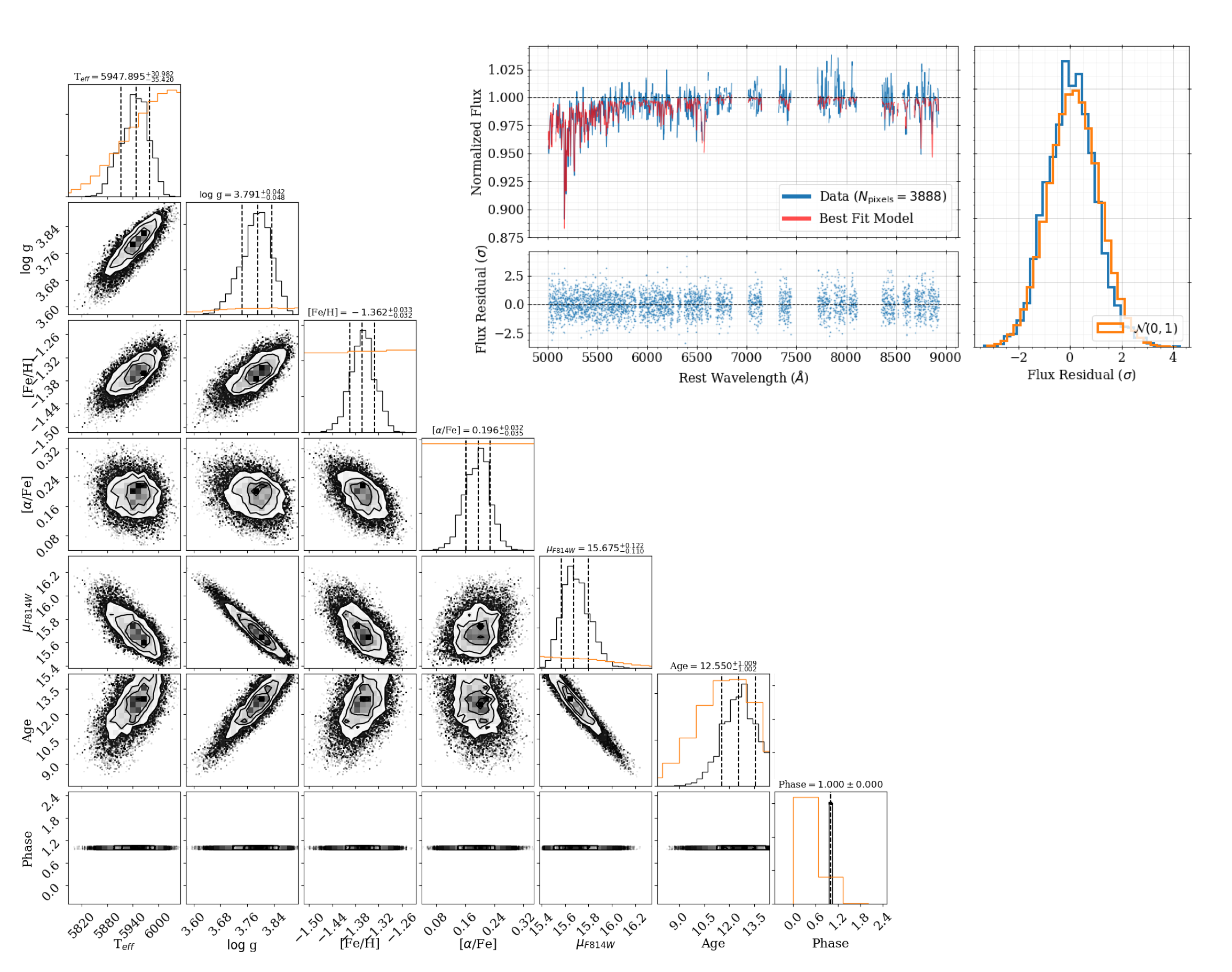}
\caption{Posterior distributions and model comparison for a star in the COSMOS field. The black points, lines, and contours in the corner plot on the left show the posterior samples, and the orange histograms show 1d projections of the prior for a given parameter. The titles above each histogram show the median and the 68\% confidence interval in that parameter. In the upper right corner, the star's spectrum shows close agreement with the best fit model (red line) as defined by the median parameters of the corner plot. As described in Section \ref{sec:methods}, various wavelength regions of the data have been masked out for the fitting process, including the central region of H$\alpha$. The uncertainty-scaled flux residuals are shown in the lower panel, and the distribution of these residuals show good agreement with the expected unit normal in the rightmost panel.}

\label{fig:halo7d_posterior_corner}
\end{center}
\end{figure*}

We use 500 MCMC walkers that are initialized with parameters drawn from the prior distributions. These walkers are updated to new parameter values at each iteration of the fitting process using the \texttt{emcee} package \citep{emcee_citation}. We generally require $\sim 500$ iterations to reach convergence, so we choose to sample for 1500 iterations for each star. We use a conservative burn-in period that throws out the first 70\% of samples, keeping the most recent 30\% as our posterior samples. Because we need to consider multiple spectral correction parameters for each observation and we consider each observation separately, this process can be quite computationally expensive in both RAM and time. An example of the pipeline's output posterior distributions and model comparison are shown in Figure \ref{fig:halo7d_posterior_corner} for a star in the COSMOS field; in this case, we see good agreement between the synthetic model and the data\footnote{The data and model in the ``Normalized Flux'' panel have been smoothed slightly for the purpose of visual comparison. The data used by the chemistry pipeline and shown in the ``Flux Residual'' spectrum and histogram are the original, unsmoothed data.}.

One limitation of our process comes from the assumption that the flux measurements at each pixel within an observation and between observations are uncorrelated, though we know this not to be true because of our preprocessing steps. While a more rigorous fitting approach would incorporate these correlations, we use the uncorrelated assumption because it simplifies the calculations and increases computation speed. 

To validate the results of our pipeline, we generate synthetic, HALO7D-like spectral observations with known stellar parameters. The results of analyzing those synthetic observations are shown in Appendix \ref{sec:fake_stars}. The main takeaway is that the abundances agree with the known parameters for stars with combined spectral SNR$> 20~\rm \AA^{-1}$\@. For SNR$ < 20~\rm \AA^{-1}$, we begin to see a bias in the abundances, so we omit any stars with combined spectral SNR below this limit for the following analyses. We also find that our posterior abundance distributions are slightly too narrow, requiring an inflation of the posterior covariance by a factor of $1.31^2$\@. This factor implies that our posterior abundance distributions have widths that are approximately 1.31 times smaller than needed to explain the disagreement with the expected values; this implies that  our systematic uncertainties in abundances are approximately 31\% of the uncertainty reported by the pipeline. All abundance uncertainties shown or reported in this paper have been inflated by this factor. Finally, we analyze MSTO stars in the well-studied globular clusters of M2 and M92 in Appendix \ref{sec:gc_comparison} and show that the pipeline is able to recover results that are consistent with the literature.

\section{Chemical Abundances of HALO7D Stars} \label{sec:results}

\begin{figure*}[t]
\begin{center}
\begin{minipage}[c]{\linewidth}
\includegraphics[width=\linewidth]{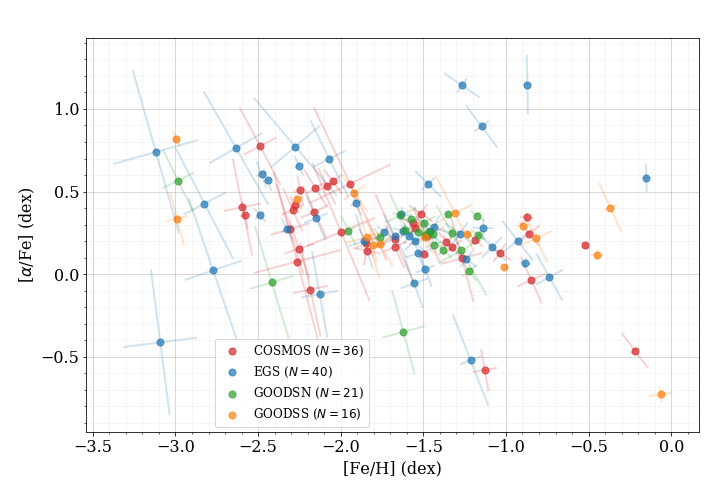}
\end{minipage}
\caption{Posterior chemical abundances of HALO7D stars colored by the field to which they belong. The errorbars are perpendicular vectors that show the 1-sigma eigenvectors of the posterior covariance matrices.}
\label{fig:HALO7D_chemistry}
\end{center}
\end{figure*}

In this section, we present our chemical abundance measurements for the HALO7D dataset. Of the 199 stars from HALO7D with measured 3D velocities (which we hereafter refer to as the ``Velocity'' sample), 113 had converged posterior results for their chemical abundances (hereafter the ``Chemistry'' sample). The numbers per field are summarized in Table~\ref{t:analysis_summary} and the abundances are displayed in Figure \ref{fig:HALO7D_chemistry}. There are several reasons why a star did not converge in the abundance pipeline, but the two main factors are (1) the resulting posterior distribution peaking too close to the edge of the model grid in at least one of the stellar parameters (i.e. $\teff$, $\logg$, $\feh$, and $\alpha$), and (2) having a combined spectral SNR$< 20~\rm \AA^{-1}$ which makes constraining the wavelength solution difficult and generally leads to unconstrained posterior distributions. The latter reason effectively acts as a magnitude cut, which makes the faint magnitude limit $m_{F606W}=23$~mag for EGS and GOODSN, 22.5~mag for COSMOS, and 22~mag for GOODSS, instead of the $m_{F606W}=24.5$~mag of the original/complete HALO7D sample. As a result, we expect that COSMOS and GOODSS cover slightly nearer distances when compared to EGS and GOODSN\@. 

\begin{table}[h]\caption{Summary of HALO7D targets in each field for different measurements. 
\label{t:analysis_summary}}
\centering
\begin{tabular}{cccc}
\hline\hline
Field        & $N_{\mathrm{v_{3D}}}$\footnote{Targets that have measured $v_{LOS}$ and proper motions from \citet{Cunningham_2019a} and \citet{Cunningham_2019b}.} & $N_{\mathrm{abundances}}$\footnote{Targets that have measured $\feh$ and $\alphafe$\@.}      \\ \hline
\multicolumn{1}{c}{COSMOS}  & 81 & 36  \\
\multicolumn{1}{c}{GOODSN}  & 32  & 21 \\
\multicolumn{1}{c}{GOODSS} & 20  & 16 \\
\multicolumn{1}{c}{EGS}      & 66 & 40  \\
\multicolumn{1}{c}{TOTAL}  & 199 & 113  \\
\hline
\end{tabular}
\end{table}

Figure \ref{fig:HALO7D_chemistry} shows the median $\feh$ and $\alphafe$ and corresponding posterior uncertainty for individual stars colored by the field they belong to. The errorbars are perpendicular lines that show the eigenvectors of the posterior covariance of the $\feh$ and $\alphafe$ distribution, with the length corresponding to 1-$\sigma$ in those eigenvectors. The bulk of stars lie in the $-2 < \feh < -1$ region, and the vast majority of stars are supersolar in $\alphafe$ as we would expect for old, halo populations and has been seen by previous studies (e.g. SDSS with \citealt{Carollo_2007,Carollo_2010}, H3 with \citealt{Conroy_2019a,Conroy_2019b,Naidu_2020}, APOGEE with \citealt{Helmi_2018,Mackereth_2019}). GOODSN (in green) has abundance distributions with smaller dispersions compared to the other fields; this is partly because GOODSN stars generally have higher average SNR spectra than the other fields (as seen in Figure \ref{fig:HALO7D_SNRs}), and thus have smaller posterior abundance distributions, though it could also be that the GOODSN field is less chemically diverse than the other fields. As we will discuss in Section \ref{sec:discussion}, the kinematics and abundances of the GOODSN stars are consistent with originating almost exclusively from the GSE progenitor, whereas the other fields appear to have multiple contributions. Excluding the clustering of GOODSN's around $\feh=-1.5$~dex, the stars of each field occupy the same regions of abundance space, and there isn't an immediately obvious difference between the fields when considering chemistry alone.

\begin{figure}[h]
\begin{center}
\includegraphics[width=\linewidth]{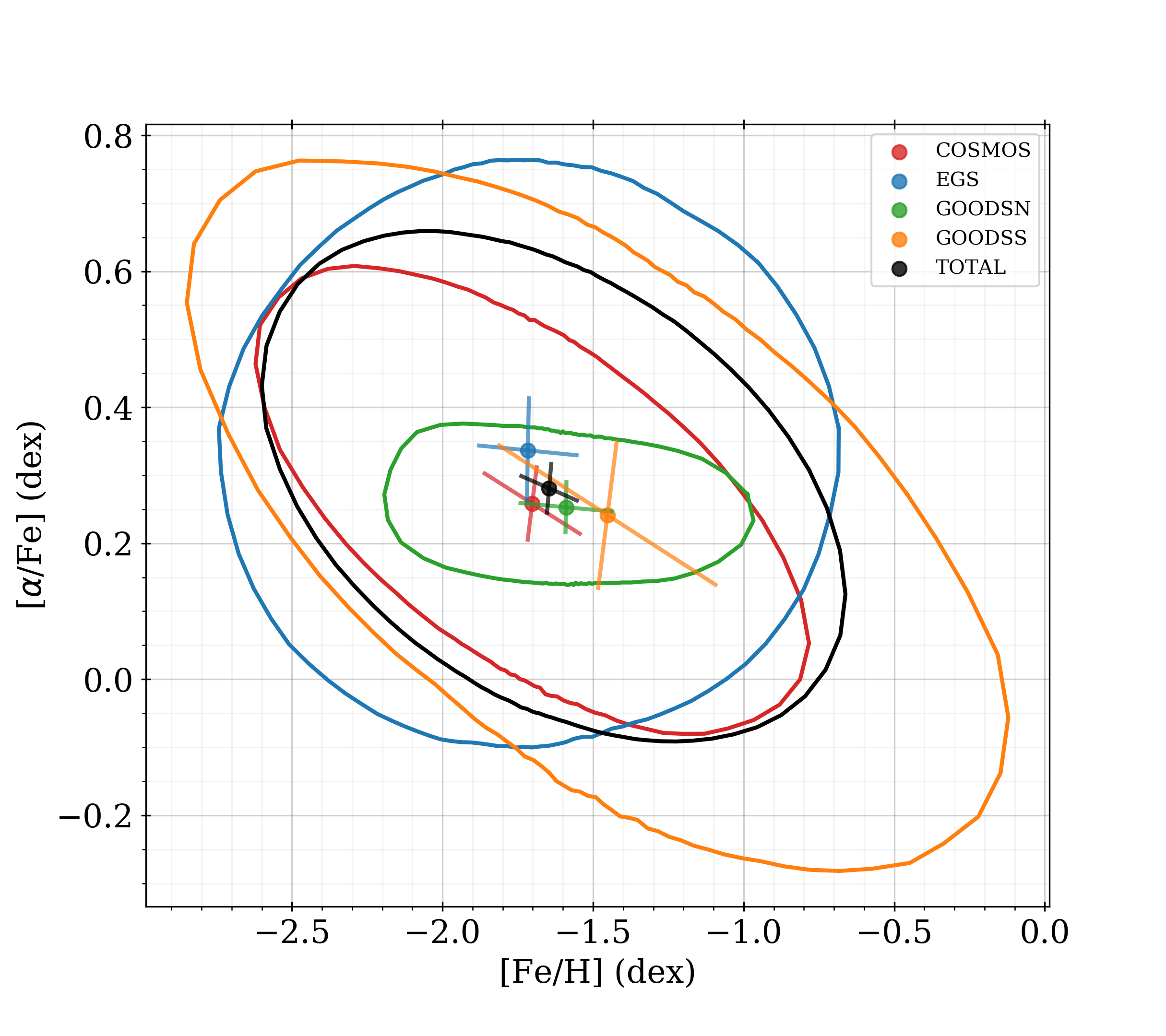}
\caption{Resulting population distribution of abundances for the HALO7D fields and total Chemistry sample after fitting 2D multivariate Gaussians. The colored data points show the median abundances, and the lines coming off of those points are perpendicular 1-sigma eigenvectors showing the uncertainty in that median. The approximately elliptical shapes show the region that the model predicts contains 68\% of the population data; these shapes are generated by taking samples of the population medians and covariance matrices from the 2D Gaussian fitting process.}
\label{fig:HALO7D_average_chemistry}
\end{center}
\end{figure}

To compare the distributions of $\feh$ and $\alphafe$ in each field more quantitatively, we fit 2D multivariate normal distributions in a hierarchical model to each field's abundances independently. The abundances in each sample are assumed to follow a 2D Gaussian distribution whose mean and covariance we are measuring, and each star's posterior abundances are assumed to be draws from that population distribution. These results are shown in Table \ref{t:HALO7D_chemistry_summary} (labelled as the ``Chemistry'' sample) and Figure \ref{fig:HALO7D_average_chemistry}. The field-colored datapoints show the median of the 2D distributions, the crosses show the 1-$\sigma$ uncertainties on the median, and the approximately elliptical shapes show the median region that the model predicts to contain 68\% of data. These 2D fits show that all the fields have similar median abundances and uncertainties. For the total sample, we find $\langle \feh \rangle = -1.65\pm0.06$~dex, $ \sigma_{\feh}  = 0.64\pm0.05$~dex, $\langle \alphafe \rangle = +0.28\pm0.03$~dex, and $\sigma_{\alphafe} = 0.24\pm0.02$~dex. As we show in Section \ref{sec:discussion}, the fraction of disk contamination in each field and for the total sample is relatively small, so the impact of foreground disk stars on these abundance distributions is minimal.

Recent studies, such as the H3 survey \citep{Conroy_2019a,Conroy_2019b}, have found a flat halo MDF (with respect to Galactocentric radius) that is peaked at $-1.2$~dex, has an approximate scatter of $0.5$~dex, and has a fairly significant tail towards low $\feh$. The H3-measured MDF is generally more metal-rich than previous studies have measured. As \citet{Conroy_2019b} point out, many of the previous studies imposed $\feh$ cuts or targeted blue stars to minimize contamination from nearby MS stars, which biases the MDF towards the metal-poor end \citep[e.g.][]{Carollo_2007, Xue_2015, Das_2016}, while H3 does not select targets based on color. The HALO7D survey also targets blue stars, which may explain why we find a more metal-poor average of $-1.65$~dex for our sample.

\begin{figure}[h]
\begin{center}
\includegraphics[width=\linewidth]{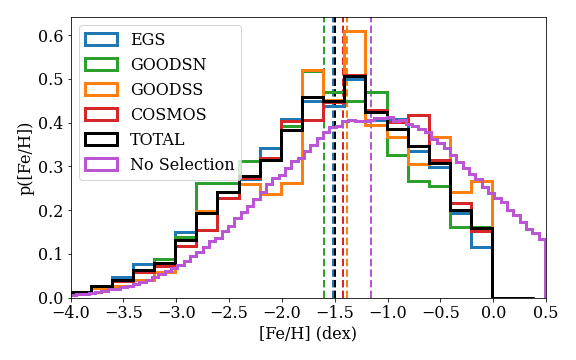}
\caption{Resulting changes to the underlying MDF (purple) when the effective selection functions in each HALO7D field are used. The vertical dashed lines show the median of each distribution. HALO7D, like other surveys that target blue stars, results in a bias towards lower $\feh$\@.}
\label{fig:selection_function}
\end{center}
\end{figure}

We estimate the impact of our selection function on the HALO7D results in Figure \ref{fig:selection_function}. For the ``No Selection'' model, we take MIST isochrones and weight them by $$\feh \sim \mathcal{N}(-1.2,1.0~\mathrm{dex})$$ $$\mathrm{age} \sim \mathcal{N}(12,2~\mathrm{Gyr})$$ $$\mathrm{mass} \sim \mathrm{Kroupa~IMF}$$ which produces the purple histogram\footnote{While our choice of $\sigma=1.0$~dex in the $\feh$ distribution is larger than the scatter measured by \citet{Conroy_2019b}, we choose to use this simplified model as a reasonable approximation that produces a significant number of low-$\feh$ stars.}. We then take those weighted isochrones and apply the observed HALO7D selection functions for each of the fields. This includes masking points outside of the model grid in Table \ref{t:model_grid}, as well as ensuring that the resulting distributions in $m_{f606W}$ and $m_{F814W}$ are exactly what is observed in the HALO7D Chemistry sample, producing the other histograms. The dashed vertical lines show the locations of each distribution's median $\feh$\@. From this, we see that the HALO7D cuts on color, magnitude, and stellar parameters bias the measured MDF towards lower $\feh$ compared to the underlying model population, and the median is shifted by $\sim -0.3$~dex. When accounting for this bias, our median $\feh$ of $-1.65\pm 0.06$~dex with a scatter of $0.64\pm 0.05$~dex is consistent with the H3 mean halo metallicity of $-1.2$~dex. 

\begin{table*}[t]
\centering
\caption{Summary of abundances for each field, and for different subsamples of the data. Measurements report the 16-th, 50-th, and 84-th percentiles. The $\feh$ bins are split at $-2.0$~dex and $-1.1$~dex based on median posterior $\feh$\@. The missing subsamples of GOODSS Low $\feh$, GOODSN High $\feh$, and GOODSN LOW $\feh$ are omitted because they have too few stars ($N \leq 3$) for a useful analysis.}
\begin{tabular}{ccccccc}
\hline \hline
Field     & Sample     & $N_{\mathrm{stars}}$                         & $\langle \feh \rangle$ & $\sigma_{\feh}$ & $\langle \alphafe \rangle$ & $\sigma_{\alphafe}$ \\ 
\hline
COSMOS & Chemistry & 36 & $-1.70_{-0.09}^{+0.11}$ & $0.61_{-0.07}^{+0.08}$ & $+0.26\pm0.04$ & $0.23_{-0.03}^{+0.04}$ \\
 & High [Fe/H] & 6 & $-0.72\pm0.15$ & $0.35_{-0.10}^{+0.22}$ & $+0.08\pm0.15$ & $0.34_{-0.10}^{+0.20}$ \\
 & Mid [Fe/H] & 16 & $-1.52_{-0.07}^{+0.06}$ & $0.24_{-0.05}^{+0.07}$ & $+0.17_{-0.07}^{+0.06}$ & $0.20_{-0.05}^{+0.07}$ \\
 & Low [Fe/H] & 14 & $-2.28\pm0.05$ & $0.17_{-0.04}^{+0.05}$ & $+0.44\pm0.04$ & $0.09_{-0.04}^{+0.05}$ \\ \hline
GOODSN & Chemistry & 21 & $-1.59_{-0.09}^{+0.10}$ & $0.42_{-0.07}^{+0.09}$ & $+0.26\pm0.02$ & $0.07\pm0.02$ \\
 & Mid [Fe/H] & 19 & $-1.48\pm0.05$ & $0.21_{-0.03}^{+0.05}$ & $+0.25\pm0.02$ & $0.07\pm0.02$ \\ \hline
GOODSS & Chemistry & 16 & $-1.43_{-0.24}^{+0.22}$ & $0.91_{-0.13}^{+0.20}$ & $+0.24\pm0.09$ & $0.35_{-0.06}^{+0.08}$ \\
 & High [Fe/H] & 6 & $-0.60\pm0.20$ & $0.47_{-0.15}^{+0.25}$ & $+0.05_{-0.21}^{+0.23}$ & $0.52_{-0.15}^{+0.32}$ \\
 & Mid [Fe/H] & 7 & $-1.61_{-0.13}^{+0.14}$ & $0.33_{-0.09}^{+0.15}$ & $+0.26_{-0.05}^{+0.06}$ & $0.08_{-0.05}^{+0.10}$ \\ \hline
EGS & Chemistry & 40 & $-1.72_{-0.11}^{+0.10}$ & $0.67\pm0.08$ & $+0.33\pm0.05$ & $0.29\pm0.04$ \\
 & High [Fe/H] & 6 & $-0.78_{-0.21}^{+0.18}$ & $0.45_{-0.14}^{+0.23}$ & $+0.33_{-0.23}^{+0.25}$ & $0.54_{-0.19}^{+0.36}$ \\
 & Mid [Fe/H] & 20 & $-1.50_{-0.05}^{+0.06}$ & $0.23_{-0.04}^{+0.05}$ & $+0.30\pm0.07$ & $0.30_{-0.05}^{+0.08}$ \\
 & Low [Fe/H] & 14 & $-2.45_{-0.09}^{+0.08}$ & $0.30_{-0.07}^{+0.10}$ & $+0.43_{-0.10}^{+0.07}$ & $0.23_{-0.08}^{+0.11}$ \\ \hline
TOTAL & Chemistry & 113 & $-1.65\pm0.06$ & $0.64_{-0.04}^{+0.05}$ & $+0.28_{-0.03}^{+0.02}$ & $0.24\pm0.02$ \\
 & High [Fe/H] & 18 & $-0.70\pm0.08$ & $0.35_{-0.06}^{+0.07}$ & $+0.15_{-0.09}^{+0.10}$ & $0.39_{-0.07}^{+0.09}$ \\
 & Mid [Fe/H] & 62 & $-1.51\pm0.03$ & $0.22\pm0.02$ & $+0.24\pm0.03$ & $0.19_{-0.02}^{+0.03}$ \\
 & Low [Fe/H] & 33 & $-2.43_{-0.06}^{+0.05}$ & $0.30_{-0.04}^{+0.05}$ & $+0.46\pm0.04$ & $0.15_{-0.03}^{+0.04}$ \\
\hline
\end{tabular}
\label{t:HALO7D_chemistry_summary}
\end{table*}

\section{Chemodynamics with HALO7D} \label{sec:discussion}

\subsection{Anisotropy Parameter, $\beta$} \label{ssec:anisotropy_calculations}
The power of a survey like HALO7D comes from being able to consider multiple dimensions together. We do this by computing field-averaged kinematics in the form of the anisotropy parameter: $$\beta = 1-\frac{\langle v_\phi^2\rangle+\langle v_\theta^2\rangle}{2\langle v_r^2\rangle},$$
\noindent as was used in the analysis of \citet{Cunningham_2019b}. With this definition, $\beta=1$ implies that a population is on entirely radial orbits, $\beta = 0$ is for a population with isotropic orbits, and $\beta \to -\infty$ for a population with entirely circular orbits. We calculate anisotropies, net halo rotation ($\langle v_\phi \rangle$), and the fraction of disk contamination ($f_{disk}$) for our four fields and the total sample using the HALO7D stars that have chemical abundance measurements. This is done by modeling the spatial densities, MDFs, age distributions, and velocity component distributions for both the disk and the halo. 

\begin{table*}[t]
\centering
\caption{Disk model distributions. For measuring kinematics from the real HALO7D data, we use the posterior $\feh$ and age distributions as measured using the chemistry pipeline. The disk's density profile distribution is chosen to be the same as used in the analysis of \citet{Cunningham_2019b}.}
\begin{tabular}{cc}
\hline \hline
Distribution                                   & Functional Form \\ \hline
\multicolumn{1}{c|}{$p(\mathrm{mass} | \feh, \mathrm{age})$}            & \citet{Kroupa_2001} IMF      \\
\multicolumn{1}{c|}{$p(\mu = m-M)$}            &  $\propto D^3 \exp\left(-\frac{R_D}{3~\mathrm{kpc}}-\frac{z}{1~\mathrm{kpc}}\right)$, where $R_D^2 = x^2+y^2$ \\
\multicolumn{1}{c|}{$p(\vec M | \feh,\mathrm{age},\mu) = \prod_f p(M_f| \feh,\mathrm{age},\mu)$}            &  $\prod_f \mathcal{N}(M_{f,\feh,\mathrm{age}} | m_f-\mu, \sigma_{m_f})$\\ \hline
\end{tabular}
\label{t:prior_disk_distributions}
\end{table*}

We first generate a ``Thick Disk'' and a ``Halo'' absolute magnitude probability distribution for each star. As in Section \ref{ssec:generating_priors}, we do this by weighting MIST isochrones using a Kroupa IMF, an age distribution, an $\feh$ distribution, each star's de-reddened color, and a density profile for either the Thick Disk or MW stellar halo. These absolute magnitude distributions give distributions on distances that we use in combination with the previously-measured PMs and LOS velocities to calculate the kinematics for each field. For both the disk and halo models, we use the posterior distributions in $\feh$ and age for each star as measured from the abundance pipeline; because of this, the isochrone weighting -- and therefore the distance distribution and corresponding velocity measurements -- for each star depends on the abundances we've measured. For the disk isochrone weighting, we use the distributions shown in Table \ref{t:prior_disk_distributions}; the density profile is chosen to match that of \citet{Cunningham_2019b}. For the halo isochrone weighting, we use the same priors as in Table \ref{t:prior_distributions}. 

\begin{table*}[t]
\centering
\caption{Velocity prior distributions for Disk and Halo model. These are the same distributions used in the analysis of \citet{Cunningham_2019b}.}
\begin{tabular}{c|c|c}
\hline \hline
Component & Distribution                                   & Functional Form \\ \hline
Halo &  $p(v_r)$ &      $\mathcal{N}(0, \sigma_r) $    \\
 &  $p(v_\phi)$ &   $\mathcal{N}(\langle v_\phi\rangle, \sigma_\phi)$ \\
 & $p(v_\theta)$  &   $ \mathcal{N}(0, \sigma_\theta)$    \\ \hline 
Disk &  $p(v_{R_D})$ &  $ \mathcal{N}(0, 45~{\rm km \ s}^{-1}) $         \\
 &  $p(v_z)$ &  $ \mathcal{N}(0, 20~{\rm km \ s}^{-1}) $ \\
 & $p(v_T)$  &  $ \mathcal{SKN}(\mu = 242~{\rm km \ s}^{-1}, \sigma=46.2~{\rm km \ s}^{-1},a=-2) $    \\ \hline 
\end{tabular}
\label{t:prior_velocity_distributions}
\end{table*}

\begin{figure*}[t]
\begin{center}
\includegraphics[trim=5cm 0 5cm 0,clip,width=\linewidth]{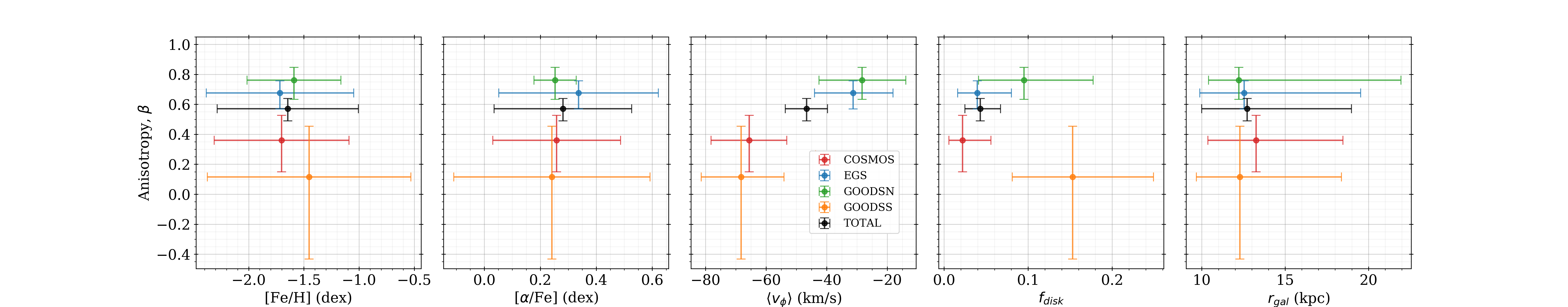}
\caption{Summaries of field distributions in $\feh$, $\alphafe$, net halo rotation $\langle v_\phi \rangle$, fraction of disk contamination $f_{disk}$, and Galactocentric radii compared to anisotropy. The abundance distributions are those fit with a 2D multivariate normal as summarized in Figure \ref{fig:HALO7D_average_chemistry} and explained in the text. The ``errorbars'' in these panels show the median and 68\% confidence interval for each distribution being considered. These plots show that the fields begin to separate when we consider kinematics. As described in the text of Section \ref{ssec:anisotropy_calculations}, we model both the disk and halo velocities component distributions, so a large $f_{disk}$ measurement does not directly cause the large prograde rotations we measure in some subsamples. EGS and GOODSN both have very radial anisotropies ($\beta \to 1$) while GOODSS is the field with the most isotropic ($\beta \sim 0$) orbits. All fields cover the same approximate radial extent. EGS and GOODSN have the smallest net rotation and the most radial anisotropies, both of which are signatures of the Gaia-Sausage-Enceladus. COSMOS and GOODSS both have large and negative net rotations and show less radial anisotropies, suggesting that these fields may have a larger contribution from the in-situ halo.}
\label{fig:HALO7D_chemistry_and_anisotropy}
\end{center}
\end{figure*}

When these absolute magnitude distributions are combined with the LOS velocities and proper motion measurements of \citet{Cunningham_2019a,Cunningham_2019b}, we have 3D velocities and positions for each star when belonging to the disk or halo, which allow us to measure the properties of the halo's velocity ellipsoid. For the velocity components, we assume the distributions shown in Table \ref{t:prior_velocity_distributions}, where $\mathcal{SKN}$ is a skew-normal distribution with shape parameter $a$\@. As before, these distributions are chosen to match those of \citet{Cunningham_2019b}. To transform between the observer frame and a Galactocentric one, we use $r_\odot = 8.5$~kpc, assume a circular speed of $235~{\rm km \ s}^{-1}$, and solar peculiar motion $(U,V,W) = (11.1,12.24,7.25)~{\rm km \ s}^{-1}$ \citep{Schonrich_2010}. Our calculations use a right-handed coordinate system such that $\langle v_\phi\rangle < 0$ corresponds to prograde motion.

We fit for the unknown halo velocity components of $\langle v_\phi \rangle$, $\sigma_{v_r}$, $\sigma_{v_\theta}$, $\sigma_{v_\phi}$, which are used to calculate the anisotropy parameter $\beta$, choosing a uniform prior on $\langle v_\phi \rangle$, and non-informative priors that are proportional to $1/\sigma_{r,\phi,\theta}$ for the dispersion components. Because each star has the possibility of belonging to the disk or the halo, we fit the population as a mixture model using $f_{disk}$, which is the fraction of contamination by foreground disk stars. We note that the velocity ellipsoid parameters (i.e. $\langle v_\phi \rangle$, $\sigma_{v_r}$, $\sigma_{v_\theta}$, $\sigma_{v_\phi}$) mentioned here and shown in the following tables and figures refer only to the halo population; we fix the parameters of the disk model velocity distributions for each LOS to those shown in the bottom of Table \ref{t:prior_disk_distributions} based on the Besan\c{c}on disk model \citep{Robin_2003}. During the fitting process, we marginalize over the absolute magnitude distributions of each star to account for uncertainties in distance. The anisotropy results for various HALO7D subsets are listed in Table \ref{t:HALO7D_anisotropy_summary}. One key takeaway is that the disk contamination is quite low in all the HALO7D fields for the chemistry sample, so we are measuring the properties of a fairly clean halo sample. The small possible number of disk stars in our sample is thus unlikely to bias the average abundances we present in Section \ref{sec:results}. 

\subsection{Anisotropy and Abundances} \label{ssec:anisotropy_versus_chemistry}

Figure \ref{fig:HALO7D_chemistry_and_anisotropy} reveals that the HALO7D fields have significant differences in their kinematics, as was first seen in \citet{Cunningham_2019b}. We note that we are considering a smaller subset of the HALO7D stars than was used in \citet{Cunningham_2019b}, so we will not directly compare our measured kinematics. With the chemistry sample of HALO7D, we find that GOODSN and EGS are more radially-biased than GOODSS and COSMOS\@.

We also notice that all of the fields have  significant net halo rotation, different  than the small or nearly zero net-rotation measured by \citet{Cunningham_2019b} and others \citep[e.g.,][]{Belokurov_2020,Bird_2021}. We also note a small apparent correlation between $f_{disk}$ and anisotropy.  These may be due to distance systematics \citep{Schonrich_2011} or our choice to model the velocity ellipsoid with a single Gaussian.  \citet{Lancaster_2019} find the halo velocity ellipsoid is well described by two components, one non-rotating and radially-biased and a second non-rotating, more isotropic population.  We use a single component as more robust choice for our smaller sample size.  As a consequence, we focus our analysis on the trends in $\langle v_{\phi}\rangle$ between different subsamples.

To assess the statistical significance of the differences we see between the fields, we assume a null hypothesis that the fields have the same halo velocity ellipsoid parameters that we find for the total population. After generating many simulations of HALO7D-like data and measuring their anisotropies, we indeed find the differences we see between the four fields are not likely to be explained by chance alone. A description and summary of these tests are presented in Appendix \ref{sec:anisotropy_pipeline}.

These anisotropy differences are also not likely explained by differences in $\beta$ as a function of average Galactocentric radius; between Galactocentric radii of 10 and 20~kpc, \citet{Loebman_2018} finds an increase in median $\beta$ from $\sim0.5$ to $\sim0.6$ using simulated MW-like galaxies. This approximate trend of $\Delta \beta / \Delta r =0.01$~kpc$^{-1}$ is not likely to explain the large anisotropy differences between the HALO7D fields. Similarly, various studies of different stellar populations in the MW stellar halo have found that the anisotropy profile does not change significantly in the 10 to 20~kpc radial range \citep[e.g.,][]{Lancaster_2019,Bird_2019,Bird_2021,Liu_2022,Wu_2022} after stars associated with Sagittarius have been removed. In these studies, however, the $\beta$ measurements come from averages over the sky whereas the HALO7D sample is able to compare along different lines of sight.

Anisotropy variations at different positions in the MW stellar halo have been previously observed. \citet{Iorio_2021}, for example, measure anisotropies of RR Lyrae in bins of Galactic Z and R (see their Figures 3 and 7). While many of their bins have high anisotropy ($\beta\sim 0.8$), there are also a handful of regions with isotropic measurements ($\beta\sim0$). It is possible that the more-isotropic HALO7D fields may intersect with some of these regions.

One possible explanation for the variation in anisotropy among the fields is differences in fractional contribution from different MW halo progenitors. For instance, the highly radial EGS and GOODSN samples could indicate that these fields are dominated by stars from the GSE -- a progenitor that is marked by stars on strongly radial orbits with no net rotation \citep[e.g.,][finds $\langle v_{\phi}\rangle\sim 0~{\rm km \ s}^{-1}$]{Belokurov_2020} -- while GOODSS and COSMOS have more significant contributions from non-GSE sources. \citet{Naidu_2020}, for instance, found that the halo fraction contributed by the GSE peaks at $r_{gal}\sim 20$~kpc and that the in-situ halo contribution becomes minimal around the same radius. Many other studies have found similar peaks in the fractional contribution from a radial halo population in the $r_{gal}$ range of 10 to 20~kpc \citep[e.g.][]{Deason_2018, Lancaster_2019, Iorio_2021, Liu_2022}. Our HALO7D fields cover the range of radii where the transition between the dominance of the in-situ halo decreases and the dominance of the GSE peaks, so differences in their average kinematics may be particularly sensitive to variation in the fractional contribution from these structures. Because all the fields have similar average chemical abundance patterns, we expect that all the fields have relatively large contributions from the GSE, and that contributions from non-GSE sources likely cover the similar regions of abundance-space, such as Wukong \citep{Naidu_2020, Yuan_2020}, Nereus \citep{Donlon_2022,Donlon_2023}, and the in-situ halo/Splash.

\subsection{Anisotropy in $\feh$ Bins} \label{ssec:anisotropy_in_bins}

\begin{figure*}[t]
\begin{center}
\includegraphics[trim=5cm 0 5cm 0,clip,width=\linewidth]{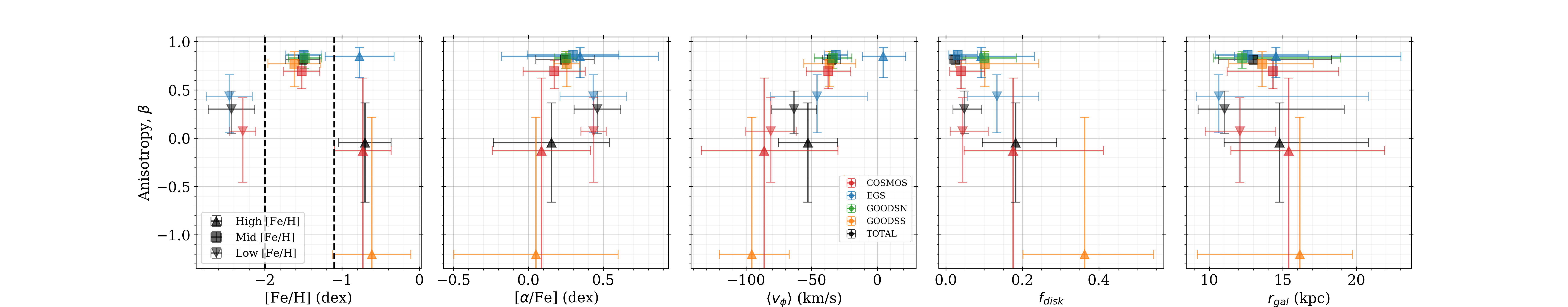}
\caption{Same as Figure \ref{fig:HALO7D_chemistry_and_anisotropy} but for different $\feh$ bins that are split at $-2.0$~dex and $-1.1$~dex. Triangles that point up correspond to the high $\feh$ bin, squares are the middle $\feh$ bin, and triangles that point down are the low $\feh$ bin. The missing $\feh$ bins for GOODSN and GOODSS are omitted for having too few stars for useful calculations. All of the mid $\feh$ bins have very radial anisotropies; the stars in this bin have metallicities and kinematics similar to the GSE. COSMOS changes in anisotropy and halo rotation more than EGS does for each of the bins. This suggests that EGS might be dominated by stars from a single progenitor while COSMOS has a higher fractional contribution from multiple progenitors. The GOODSS high $\feh$ bin has the most circular orbits compared to the other fields, suggesting that this subsample may contain the largest fraction of in-situ halo stars.}
\label{fig:FeH_binned_chemistry_and_anisotropy}
\end{center}
\end{figure*}

To explore the relationship between kinematics and chemistry, we split the sample up into different $\feh$ bins. Based on previous MW halo inventory studies, the high $\feh$ bin ($\feh > -1.1$~dex) is where we expect the largest fraction of in-situ halo/disk stars and the mid $\feh$ bin ($-2 < \feh < -1.1$~dex) is where we expect stars from the GSE to dominate. The binned abundances and kinematics are summarized in Tables \ref{t:HALO7D_chemistry_summary} and \ref{t:HALO7D_anisotropy_summary} and Figure \ref{fig:FeH_binned_chemistry_and_anisotropy}. 

All fields in the mid $\feh$ bin have highly radial anisotropies and a small net halo rotation ($\langle v_\phi \rangle$), which suggests that stars in this bin are chemodynamically similar and thus may originate from a single progenitor. In the low and high $\feh$ bins, we notice differences in the kinematics between the fields. In particular, the high $\feh$ bin shows the largest differences in anisotropy between the fields, with EGS having the most radial stars and GOODSS having the most circular stars. As before, we test the probability of observing these results by chance alone and find that the anisotropy differences in the high $\feh$ bin are likely statistically significant; these tests are summarized in Appendix \ref{sec:anisotropy_pipeline}. These anisotropy differences between the fields may be caused by variations in the fractional contributions of different progenitors, which changes the average kinematics we measure. These differences highlight the additional information available on smaller spatial scales that a survey like HALO7D is able to capture.

The mid $\feh$ bin covers $-2 < \feh < -1.1$~dex in metallicity and spans Galactocentric radii of $\sim 10 - 20$~kpc in all fields. This is the radial and metallicity range that other studies \citep[e.g.][]{Naidu_2020} have measured the GSE to dominate the halo. These facts, combined with the low net rotation and highly radial anisotropy we measure in the mid $\feh$ bins, are evidence that all the fields have a dominant fractional contribution from the GSE in the $-2 < \feh < -1.1$~dex range. Because we measure a non-zero $\langle v_\phi \rangle$ of $\sim -30~{\rm km \ s}^{-1}$ in this bin, GSE is almost certainly not the only progenitor present in this bin; however, the relative contributions from prograde structures must be relatively consistent between the different fields to produce a similar net rotation. The high $\feh$ bin has the largest range of anisotropies and $v_\phi$ values compared to the other bins. This is the bin that we expect to contain a significant number of stars from the GSE, the in-situ halo, and a few other known MW progenitors because its higher metallicity intersects with the MDFs of multiple structures. As a consequence, this large number of progenitor options is likely causing the increased variation we see in kinematics in this bin. Like the mid $\feh$ bin, the low $\feh$ bin doesn't show much variation between the chemodynamics of the fields.  

The results from EGS and COSMOS are particularly useful to compare because they have approximately the same number of stars in each of the $\feh$ bins, which limits the impact that sample size has on these comparisons. EGS has anisotropies in each $\feh$ bin that are quite radial and fairly consistent with each other, which suggests that EGS is dominated by stars from a single progenitor in all $\feh$ bins. As explained in the previous paragraph, the origin of these stars may be the GSE, as the EGS stars show the characteristic radial bias and low net rotation previously measured for the GSE. For COSMOS, the anisotropy and net halo rotation change more noticeably between the $\feh$ bins. At high and low $\feh$, the COSMOS anisotropies are isotropic and the net rotations are prograde, which suggests that these bins have significant contributions from sources other than that which produced the kinematics in the mid $\feh$ bin. Like the other fields, the COSMOS mid $\feh$ bin has kinematics and chemistry that are consistent with the GSE\@. A possible origin for the high-$\feh$, prograde stars in COSMOS is the in-situ halo, while the low-$\feh$, prograde stars may originate from the Wukong, Nereus, and/or the metal-poor tail of the in-situ halo. Because COSMOS is closer to the disk plane than any of the other HALO7D fields and EGS is the farthest, the larger contribution of in-situ halo stars in COSMOS and a smaller contribution in EGS is not unexpected. All the HALO7D fields are located far away from any Sagittarius Stream debris, so this is not likely to explain the difference in chemodynamics we observe between the fields. 

In summary, the HALO7D fields show significant differences when considering chemistry and kinematics together. Looking at different $\feh$ bins reveals that all the fields may have a significant fraction of stars from the GSE in the middle $\feh$ bin of $-2 < \feh < -1.1$~dex. EGS has similar kinematic properties in its high and low $\feh$ bins, and these are again consistent with debris from GSE\@. COSMOS, on the other hand, has more variation in its kinematics across the $\feh$ bins which suggests it has larger fractional contributions from non-GSE sources. The high $\feh$ bin has properties consistent with a large fraction of in-situ halo stars, the mid $\feh$ bin has properties like the GSE, and the low $\feh$ bin has properties like the metal-poor tail of the in-situ halo, the Wukong progenitor, and/or the Nereus progenitor. This spatial non-uniformity is compelling evidence that the MW stellar halo is not uniformly mixed in its chemodynamical distributions. 

\section{Summary} \label{sec:conclusion}

We have measured $\feh$ and $\alphafe$ abundances for 113 main sequence turn-off MW stellar halo stars across four CANDELS fields in the HALO7D survey. By focusing on MSTO stars in the stellar halo at Galactocentric radii in the range $10-40$~kpc, HALO7D is able to measure MW halo properties on smaller spatial scales than other contemporary surveys. Our abundances are combined with the previously-measured 3D positions and 3D velocities from HALO7D \citep{Cunningham_2019a, Cunningham_2019b} to measure the variation in average chemodynamical properties along each LOS; these properties include the net halo rotation, anisotropy, and average abundances. To measure our abundances for HALO7D, we have created a Bayesian pipeline that uses photometric and spectroscopic information to determine stellar parameters ($\teff,\, \logg$, age, and distance modulus) and chemical composition ($\feh,\, \alphafe$) for MSTO stars without known distances (Section \ref{sec:methods}). Our key results include:

\begin{enumerate}
    \item The abundance patterns in each of the HALO7D fields agree with each other. The average $\feh$ of the full 113 star HALO7D Chemistry sample is $-1.65$~dex with a scatter of $0.61$~dex, which is more metal-poor than some recent contemporary surveys \citep[e.g., $\langle \feh \rangle \sim -1.2$~dex for][]{Conroy_2019b,Naidu_2020}, but this is almost certainly because of our blue selection function. The average $\alphafe$ for the HALO7D Chemistry sample is $+0.28$~dex with a scatter of $0.24$~dex, which is in agreement with what we would expect for a sample drawn from a population of old, metal-poor halo stars. (Section \ref{sec:results}, Figures \ref{fig:HALO7D_chemistry}, \ref{fig:HALO7D_average_chemistry})
    \item The HALO7D fields separate in kinematic-space when we measure average properties like the anisotropy parameter $\beta$\@. EGS and GOODSN show more radially-biased orbits and near-zero halo rotation compared to GOODSS and COSMOS which have more isotropic orbits and fairly negative halo rotation. (Section \ref{ssec:anisotropy_versus_chemistry}, Figure \ref{fig:HALO7D_chemistry_and_anisotropy})
    \item Breaking the HALO7D fields into low, mid, and high $\feh$ bins at $-2.0$~dex and $-1.1$~dex shows differences in the chemodynamic makeup of the fields. All the fields have similar anisotropy in the mid $\feh$ bin, but the high and low $\feh$ bins show differences between the fields. EGS has relatively similar anisotropy and net halo rotation between the $\feh$ bins,  all of which are similar to the properties of Gaia-Sausage-Encaladus. COSMOS, on the other hand, has a mid $\feh$ bin with kinematics that are similar to the GSE, but its low and high $\feh$ bins have kinematics that suggests these bins have larger contributions from prograde structures, such as the kicked-up disk/in-situ halo, the metal-poor progenitors of Nereus and Wukong. These chemodynamical differences between the fields, even at the same Galactocentric radii, suggest that the MW stellar halo is not uniformly mixed along different lines of sight.  (Section \ref{ssec:anisotropy_in_bins}, Figure \ref{fig:FeH_binned_chemistry_and_anisotropy})
\end{enumerate}

Future work will focus on studying the full 7D chemodynamic relationship of our stars, such as using a MW potential model to estimate individual orbits, which will allow us to better quantify the relative contributions from different progenitors in each field. We are also in the process of expanding the HALO7D survey to include more LOS and more stars along each LOS, which will allow us to measure how the chemodynamical distributions change as a function of 3D position with the goal of contributing to a more complete picture of our Galaxy's accretion history.

\acknowledgments

The authors thank the anonymous referee for comments that helped improve the clarity of this paper. KM, CMR, PG were supported by NSF Grant AST-1616540 and by NASA through grant HST-AR-16625.005-A awarded by the Space Telescope Science Institute. ECC acknowledges support for this work provided by NASA through the NASA Hubble Fellowship Program grant HST-HF2-51502.001-A awarded by the Space Telescope Science Institute, which is operated by the Association of Universities for Research in Astronomy, Inc., for NASA, under contract NAS5-26555.  IE acknowledges generous support from a Carnegie-Princeton Fellowship through Princeton University.  ENK acknowledges support from the National Science Foundation under Grant No.\ AST-2233781.

The authors recognize and acknowledge the very significant cultural role and reverence that the summit of Maunakea has always had within the indigenous Hawaiian community. We are most fortunate to have the opportunity to conduct observations from this mountain.

KM thanks Avesta Rastan for helpful conversations that improved the presentation of scientific figures. KM also thanks Brian DiGiorgio for insightful conversations that led to significant reductions in computation time. 

\facilities{HST, Keck II (DEIMOS)}

\software{Astropy \citep{astropy_2013,astropy_2018,astropy_2022}, corner \citep{corner_citation}, dustmaps \citep{dustmaps_citation}, emcee \citep{emcee_citation},  IPython \citep{ipython_citation}, jupyter \citep{jupyter_citation}, matplotlib \citep{matplotlib_citation}, numpy \citep{numpy_citation}, scipy \citep{scipy_2020a,scipy_2020b}, scikit-learn \citep{scikit-learn_citation}, spec2d \citep{Cooper_2012}}

\movetabledown=30mm
\begin{rotatetable*}
\begin{deluxetable*}{ccccccccccccc}
\tablecaption{Summary of anisotropy, distance, and Galactocentric radius for each field. Measurements report the 16-th, 50-th, and 84-th percentiles, and we use the convention that $v_\phi < 0$ corresponds to prograde rotation. The subsamples are the same as listed in Table \ref{t:HALO7D_chemistry_summary}. \label{t:HALO7D_anisotropy_summary}}
\startdata
Field     & Sample                              & $\langle v_\phi \rangle$& $\sigma_{\phi}$ & $\sigma_\theta$ & $\sigma_{r}$ & $\langle D \rangle$ & $D$ Range  & $\langle r \rangle$ & $r$ Range  & $f_{disk}$ & $\beta$ \\ 
     &      & (${\rm km \ s}^{-1}$) & (${\rm km \ s}^{-1}$) & (${\rm km \ s}^{-1}$) & (${\rm km \ s}^{-1}$) & (kpc) & (kpc) & (kpc) & (kpc)  &  &  & \\ \hline
COSMOS & Chemistry & $-66_{-13}^{+12}$ & $71_{-8}^{+10}$ & $69_{-7}^{+9}$ & $106_{-12}^{+14}$ & $7.3\pm0.3$ & $3.4 - 13.3$ & $13.3_{-0.2}^{+0.3}$ & $10.4 - 18.5$ & $0.02_{-0.02}^{+0.03}$ & $+0.36_{-0.21}^{+0.17}$ \\
 & High [Fe/H] & $-86_{-48}^{+56}$ & $96_{-31}^{+57}$ & $44_{-16}^{+22}$ & $98_{-26}^{+49}$ & $9.6_{-0.7}^{+1.2}$ & $5.0 - 17.1$ & $15.2_{-0.6}^{+1.1}$ & $11.5 - 21.9$ & $0.18_{-0.13}^{+0.24}$ & $-0.13_{-1.73}^{+0.75}$ \\
 & Mid [Fe/H] & $-37\pm17$ & $58_{-10}^{+14}$ & $54_{-9}^{+11}$ & $116_{-18}^{+25}$ & $8.6_{-0.6}^{+0.8}$ & $4.6 - 13.7$ & $14.4_{-0.5}^{+0.6}$ & $11.2 - 18.8$ & $0.04_{-0.03}^{+0.06}$ & $+0.69_{-0.18}^{+0.12}$ \\
 & Low [Fe/H] & $-81\pm19$ & $71_{-13}^{+18}$ & $91_{-15}^{+22}$ & $106_{-17}^{+23}$ & $5.7_{-0.2}^{+0.3}$ & $2.4 - 8.8$ & $12.0\pm0.2$ & $9.7 - 14.5$ & $0.04_{-0.03}^{+0.07}$ & $+0.07_{-0.53}^{+0.35}$ \\ \hline
GOODSN & Chemistry & $-28\pm14$ & $55_{-10}^{+12}$ & $75_{-13}^{+16}$ & $144_{-21}^{+27}$ & $6.4_{-0.3}^{+0.4}$ & $3.8 - 17.6$ & $12.2_{-0.2}^{+0.3}$ & $10.4 - 22.0$ & $0.10_{-0.05}^{+0.08}$ & $+0.76_{-0.13}^{+0.09}$ \\
 & Mid [Fe/H] & $-34\pm14$ & $53_{-10}^{+13}$ & $59_{-13}^{+17}$ & $152_{-22}^{+31}$ & $6.4_{-0.3}^{+0.4}$ & $3.6 - 14.3$ & $12.2_{-0.2}^{+0.3}$ & $10.3 - 18.9$ & $0.10_{-0.06}^{+0.08}$ & $+0.83_{-0.11}^{+0.07}$ \\ \hline
GOODSS & Chemistry & $-68_{-13}^{+14}$ & $47_{-9}^{+13}$ & $136_{-23}^{+31}$ & $121_{-20}^{+28}$ & $6.0_{-0.3}^{+0.4}$ & $2.3 - 13.1$ & $12.3_{-0.2}^{+0.3}$ & $9.7 - 18.4$ & $0.15_{-0.07}^{+0.10}$ & $+0.11_{-0.55}^{+0.34}$ \\
 & High [Fe/H] & $-96_{-25}^{+28}$ & $46_{-19}^{+38}$ & $223_{-68}^{+126}$ & $124_{-36}^{+64}$ & $8.3\pm0.5$ & $1.4 - 14.6$ & $14.4\pm0.4$ & $9.2 - 19.7$ & $0.36_{-0.16}^{+0.18}$ & $-1.20_{-3.63}^{+1.42}$ \\
 & Mid [Fe/H] & $-36\pm20$ & $42_{-14}^{+20}$ & $55_{-13}^{+21}$ & $124_{-28}^{+44}$ & $7.5_{-0.5}^{+0.8}$ & $4.7 - 11.6$ & $13.5_{-0.4}^{+0.6}$ & $11.3 - 17.1$ & $0.10_{-0.08}^{+0.14}$ & $+0.77_{-0.24}^{+0.12}$ \\ \hline
EGS & Chemistry & $-31\pm13$ & $78_{-9}^{+10}$ & $69_{-8}^{+9}$ & $137_{-14}^{+17}$ & $8.7\pm0.3$ & $4.6 - 17.1$ & $12.5_{-0.2}^{+0.3}$ & $9.9 - 19.5$ & $0.04_{-0.02}^{+0.04}$ & $+0.68_{-0.11}^{+0.08}$ \\
 & High [Fe/H] & $+5_{-16}^{+17}$ & $35_{-14}^{+20}$ & $35_{-10}^{+16}$ & $99_{-24}^{+40}$ & $11.3_{-0.9}^{+1.0}$ & $7.6 - 20.9$ & $14.6\pm0.8$ & $11.7 - 23.0$ & $0.09_{-0.07}^{+0.14}$ & $+0.85_{-0.22}^{+0.09}$ \\
 & Mid [Fe/H] & $-32\pm9$ & $34_{-7}^{+9}$ & $49_{-8}^{+11}$ & $132_{-19}^{+24}$ & $8.8_{-0.5}^{+0.4}$ & $5.6 - 13.9$ & $12.6\pm0.3$ & $10.4 - 16.7$ & $0.03_{-0.02}^{+0.05}$ & $+0.86_{-0.07}^{+0.05}$ \\
 & Low [Fe/H] & $-46_{-35}^{+38}$ & $126_{-23}^{+32}$ & $102_{-19}^{+26}$ & $165_{-28}^{+40}$ & $5.7_{-0.2}^{+0.3}$ & $2.9 - 18.5$ & $10.5_{-0.1}^{+0.2}$ & $9.1 - 20.8$ & $0.13_{-0.08}^{+0.11}$ & $+0.43_{-0.38}^{+0.23}$ \\ \hline
TOTAL & Chemistry & $-47\pm7$ & $70_{-5}^{+6}$ & $80_{-5}^{+6}$ & $125_{-8}^{+9}$ & $7.6_{-0.2}^{+0.3}$ & $3.5 - 14.8$ & $12.7\pm0.2$ & $10.0 - 19.0$ & $0.04\pm0.02$ & $+0.57_{-0.08}^{+0.07}$ \\
 & High [Fe/H] & $-53_{-22}^{+23}$ & $78_{-15}^{+20}$ & $106_{-18}^{+25}$ & $101_{-16}^{+23}$ & $10.5_{-0.7}^{+1.0}$ & $4.6 - 16.1$ & $14.9_{-0.7}^{+1.0}$ & $11.0 - 20.8$ & $0.18_{-0.09}^{+0.11}$ & $-0.05_{-0.62}^{+0.41}$ \\
 & Mid [Fe/H] & $-35\pm7$ & $47\pm5$ & $52_{-5}^{+6}$ & $130_{-11}^{+13}$ & $8.2_{-0.3}^{+0.4}$ & $4.8 - 13.7$ & $13.0_{-0.2}^{+0.3}$ & $10.6 - 18.3$ & $0.02_{-0.02}^{+0.03}$ & $+0.82_{-0.05}^{+0.04}$ \\
 & Low [Fe/H] & $-63\pm17$ & $94_{-11}^{+14}$ & $104_{-12}^{+14}$ & $132_{-15}^{+19}$ & $5.0_{-0.2}^{+0.3}$ & $2.6 - 16.4$ & $11.1\pm0.3$ & $9.2 - 19.2$ & $0.05_{-0.03}^{+0.05}$ & $+0.30_{-0.25}^{+0.19}$ \\
\enddata
\end{deluxetable*}
\end{rotatetable*}

\bibliographystyle{aasjournal}
\bibliography{referee_revised_arxiv.bbl}

\appendix

\section{Chemical Abundance Pipeline Internal Error with Fake Stars} \label{sec:fake_stars}

We test the ability of our pipeline to recover input chemical abundances across a range of stellar parameters and spectral SNR by generating fake stars with known photometry and stellar parameters. We create two types of fake stars; those with single spectral observations that model our real observations of globular cluster stars in M2 and M92 (see Appendix \ref{sec:gc_comparison}), and those with multiple spectroscopic visits that model real HALO7D data. In each case, a fake star consists of $(m_
{F606W},m_{F814W})$ apparent magnitudes and between one and ten separate spectral observations at a chosen SNR\@. 

In the case of the globular cluster analog fake stars, we use the literature-defined MIST isochrone to define the photometry and stellar parameters. We do this by taking a real globular cluster star, and finding the closest point on the MIST isochrone in terms of photometry; the fake star then inherits that point's absolute magnitudes ($M_{F606W},~M_{F814W}$) and stellar parameters ($\teff,~\logg$) as well as the cluster distance, age, and $\feh$\@. The $\alphafe$ is randomly drawn from a uniform distribution in the range of $-0.2$~dex to $+0.6$~dex. With the fake star's stellar parameters and abundances defined, we can smooth the corresponding synthetic model spectrum by the real globular cluster star's observed seeing and LSF, and then re-bin onto the observed DEIMOS wavelength array. We next multiply the smoothed, continuum-normalized fake spectrum by an estimate of the real star's continuum (as described in Section \ref{sssec:continuum_norm}) to obtain DEIMOS-like observations for the fake star. Finally, the chosen SNR of the fake star's spectrum is used to apply flux noise and to define the flux uncertainty in each pixel. With the fake star having apparent magnitudes and a spectral observation at a chosen SNR, we are able to feed it to the chemical abundance pipeline by following the process listed in Section \ref{sec:methods}. 

For the HALO7D-like fake star analogs, we follow a similar process as with the globular cluster stars. To define the input parameters of the fake star, we use the real HALO7D star's posterior distribution from its analysis using the chemical abundance pipeline; the fake star's parameters are chosen to be the median of the posterior distribution, and the apparent magnitudes are exactly those of the HALO7D star. The corresponding synthetic model spectrum is repeated so that the fake star is given the same number of spectroscopic visits/observations as the real HALO7D star that it is based on. The seeing, LSF, and continuum of each real spectroscopic observation are used to define the smoothing and continuum of the different fake spectral visits. Next, a chosen SNR defines the resulting combined SNR of the different spectroscopic visits, where the SNRs of the individual visits are chosen to follow the same ratio of the real HALO7D star; the SNR of each spectroscopic visit sets the flux noise and the flux uncertainty in each pixel. As with HALO7D stars, the apparent magnitudes and multiple spectroscopic visits of the fake stars are processed through the chemistry pipeline. 

\begin{figure*}[t]
\begin{center}
\includegraphics[width=\linewidth]{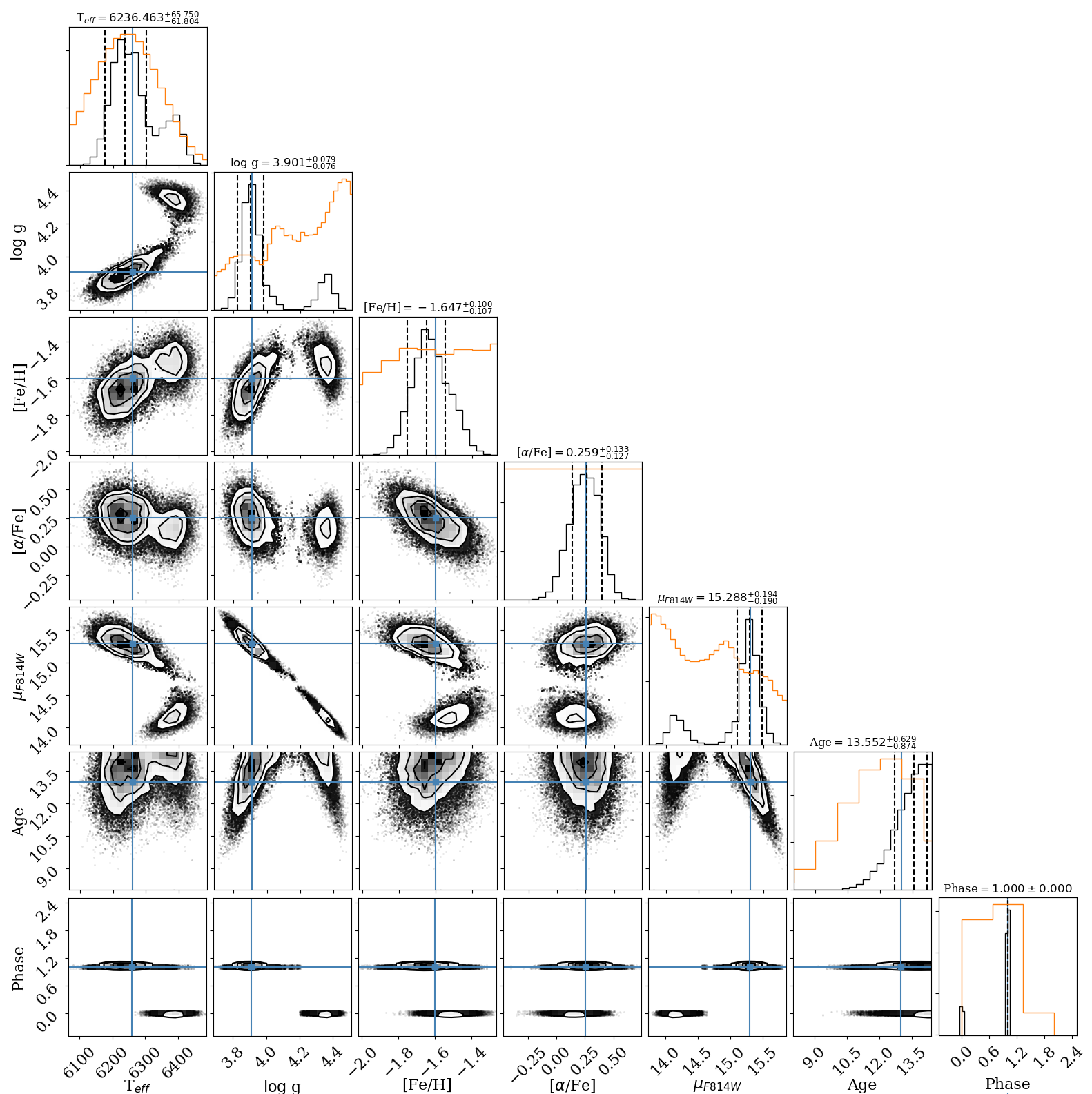}
\caption{Posterior distributions for a fake star that has a single spectral observation with median SNR of $40~\rm \AA^{-1}$ with stellar parameters that are near the MSTO of M2\@. The blue lines show the location of the fake star's true parameters, the black points/lines/contours show the posterior samples, and the orange histograms show 1d projections of the prior for a given parameter. The titles above each histogram show the median and the 68\% confidence interval in that parameter.}
\label{fig:fakestar_corners}
\end{center}
\end{figure*}

For each real star that defines a set of stellar parameters and abundances, we create fake stars with varying SNR from $5~\rm \AA^{-1}$ to $200~\rm \AA^{-1}$\@. These fake stars are processed identically to the real stars following the methods of Section \ref{sec:methods} to assess the pipeline's ability to recover the input parameters as a function of spectral SNR. An example posterior distribution for one fake star with spectral SNR of 40$~\rm \AA^{-1}$ and properties that place it near the MSTO of M2 (i.e. distance of 11~kpc, age of 13~Gyr, [Fe/H] of $-1.65$~dex) is shown in Figure \ref{fig:fakestar_corners}. The black points and lines show the posterior samples that have good agreement with the blue lines which show the values of the input parameters. In cases where the pipeline incorrectly measures $\logg$ -- that is, when the star is assigned the incorrect phase, placing it on one branch when it belongs to another -- we find that the chemistry is unchanged; specifically, the posterior $\feh$ and $\alphafe$ samples are in agreement for each of the possible phases. Since we are most interested in measuring abundances, we include stars with incorrectly-measured posterior phase/$\logg$. 

\begin{figure}[h]
\begin{center}
\begin{minipage}[c]{1.0\linewidth}
\includegraphics[width=\linewidth]{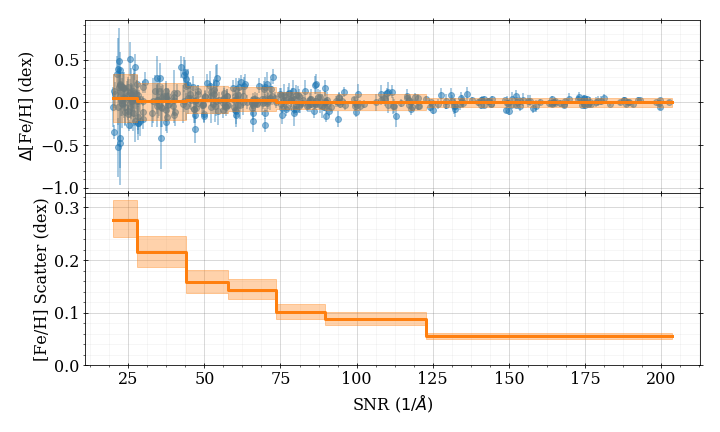}
\end{minipage}\\
\begin{minipage}[c]{1.0\linewidth}
\includegraphics[width=\linewidth]{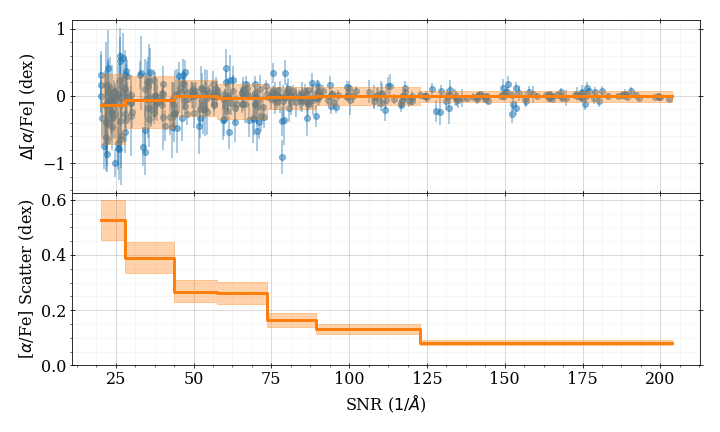}
\end{minipage}
\caption{Comparison of measured posterior $\feh$, $\alphafe$ and input truth for fake spectra as a function of spectral signal-to-noise. Blue errorbar points show the 68\% confidence interval of the posterior around the median. The orange lines in the top panels show the median and 68\% scatter within SNR bins of 50 stars and the orange lines in the bottom panels show the median 68\% width of the posterior samples (and corresponding 1-sigma uncertainty) in those same SNR bins. We remove observations with spectral SNR $< 20 ~\rm \AA^{-1}$ because the abundance uncertainties begin to show systematic biases at this level.}
\label{fig:fakestar_chemistry_snr}
\end{center}
\end{figure}

The posterior chemical abundance distributions of the fake stars are then used to assess the internal errors of our chemistry pipeline and the reasonableness of the posterior uncertainties. The 2D abundance distance between the input truth and the posterior median is measured using the covariance of the posterior $\feh$ and $\alphafe$ samples; these distances should follow a chi-squared distribution with 2 degrees of freedom (DOF) and a scale of 1 if the posterior uncertainties are reasonable. While the distribution we measure does indeed follow a chi-square with 2 DOF, we instead measure a scale factor of 1.31, meaning that our posterior widths need to be inflated by this amount (i.e. the covariance matrix needs to be inflated by $1.31^2$) to capture the pipeline's true uncertainty. All remaining stellar parameter measurements in this paper, including Figure~\ref{fig:fakestar_chemistry_snr} and all other figures, have had their posterior uncertainties increased by this factor. After applying this inflation, we still noticed a slight bias in the $\feh$ and $\alphafe$ results for fake stars with spectral SNR $< 20~\rm \AA^{-1}$\@. As a result, we ignored real and fake stars with spectral SNR $< 20~\rm \AA^{-1}$ from all analyses. 

\begin{figure}[h]
\begin{center}
\begin{minipage}[c]{\linewidth}
\includegraphics[width=\linewidth]{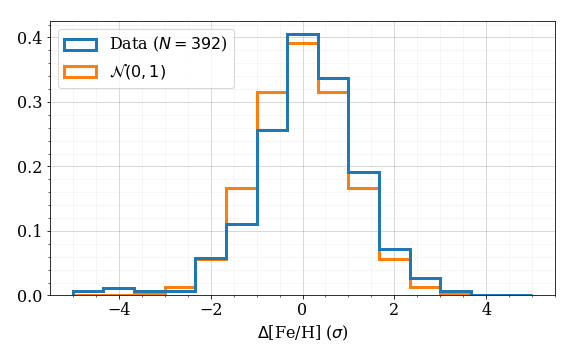}
\end{minipage}\\
\begin{minipage}[c]{\linewidth}
\includegraphics[width=\linewidth]{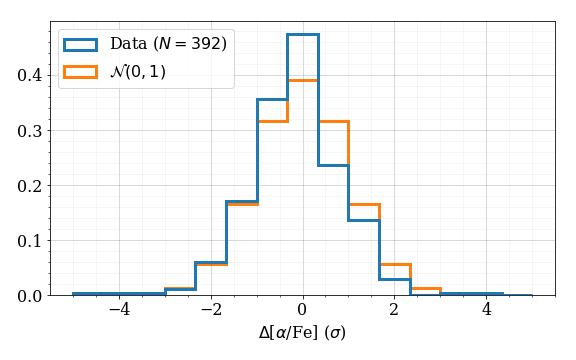}
\end{minipage}
\caption{Uncertainty-scaled comparison of measured posterior $\feh$, $\alphafe$ and input truth for fake spectra. The unit Gaussian (orange line) shows that the pipeline is returning reasonable abundances and corresponding uncertainties.}
\label{fig:fakestar_chemistry_hists}
\end{center}
\end{figure}

Plotting the difference between the posterior abundances and the input truth versus input spectral SNR, as in Figure \ref{fig:fakestar_chemistry_snr}, shows that the pipeline recovers useful (posterior disagreement $< 0.25$~dex) $\feh$ above SNR $\sim 25~\rm \AA^{-1}$ and $\alphafe$ above SNR $\sim 45~\rm \AA^{-1}$\@. This is similar performance to other spectroscopic analysis pipelines of DEIMOS data \citep[e.g.,][]{Escala_2019,Kirby_2008}. The uncertainty-scaled difference between truth and posterior abundance is shown in Figure~\ref{fig:fakestar_chemistry_hists}. These distributions show good agreement with the unit Gaussian (shown in orange), implying that the posterior abundances and uncertainties are reasonable. Our choice of a uniform prior on $\alphafe$ that has no correlation with the other stellar parameters can be improved in the future with access to MIST isochrones that contain different values of $\alphafe$; this flat $\alphafe$ prior currently plays a part in causing the relatively large scatter on our ability to recover $\alphafe$ as compared to $\feh$. 

Our posterior uncertainties are relatively large compared to other pipelines because of our large prior uncertainties in $\logg$, which come from unknown distances. Many other investigations that measure stellar abundances are able to use previously-measured parallaxes or distances to tightly constrain the possible $\logg$ values, which helps to return better-measured abundances from the same spectral SNR and to push the limiting SNR to much lower values. Our necessarily diffuse distance priors cause the pipeline to consider a wider range of models and thus return more uncertain abundances.  

While a given star's posterior $\feh$ and $\alphafe$ measurements are correlated (as seen in Figure~\ref{fig:fakestar_corners}, as well as our other posterior distributions), we want to ensure that the pipeline doesn't produce a correlation in any population statistics (e.g. group means and spreads) measured from a collection of stars. To test this, we fit a hierarchical model using the $\Delta\feh$ and $\Delta \alphafe$\footnote{Here, $\Delta X = X_{\mathrm{posterior}}-X_{\mathrm{input}}$} posterior distributions of all fake stars to measure a pooled population mean and covariance matrix of a 2D Multivariate Normal. The posterior mean of the population distribution was centered at $(\Delta\feh,\Delta\alphafe) = (0,0)$~dex, corresponding spreads of $(0.02, 0.01)$~dex, and a correlation between $\Delta\feh$ and $\Delta\alphafe$ that is consistent with 0\@. Finally, we also perform a prior sensitivity analysis by changing the parameters of the $\feh$ and age distributions in Table \ref{t:prior_distributions} while re-analyzing the fake stars; we find that the posterior chemical abundances are largely unchanged for fake stars with spectral $\mathrm{SNR} > 20~\rm \AA^{-1}$ when using reasonable choices of the $\feh$ and age distribution parameters. Therefore, the pipeline does not induce artificial correlations in the population statistics.

\section{Chemical Abundance Pipeline External Validation with Globular Clusters} \label{sec:gc_comparison}

\begin{figure}[h]
\begin{center}
\includegraphics[width=\linewidth]{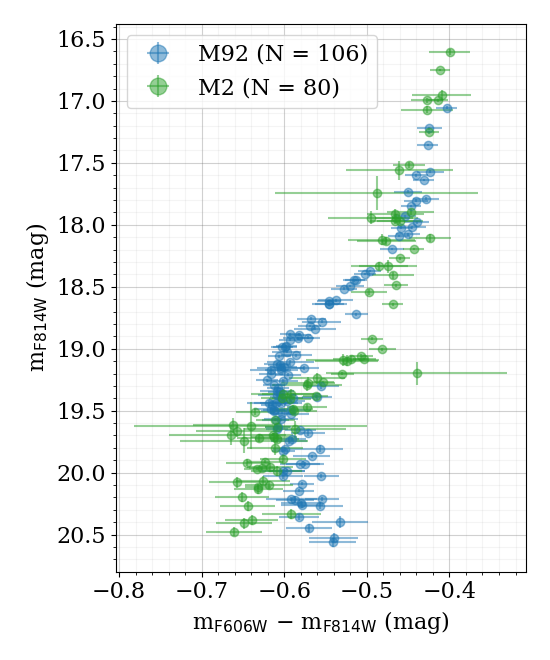}
\caption{Color-Magnitude Diagram of the M2 and M92 globular cluster stars in HST filters (STMAG) used in our validation sample. The HST photometry was calculated by transforming ground-based photometry, which is why the uncertainties are quite large. Both clusters have stars that are around the MSTO\@.}
\label{fig:GC_CMDs}
\end{center}
\end{figure}

In Appendix \ref{sec:fake_stars}, we used fake stars with well-behaved data to test our chemical abundance pipeline. In this appendix, we assess the pipeline's ability to recover chemical abundances using real data from well-studied globular clusters: M2 and M92\@. M2 is at a heliocentric distance of 11~kpc and has an iron abundance of $-1.65$~dex \citep[][2010 edition]{Harris_1996}, which is relatively typical of MW halo stars. M92 is at a heliocentric distance of 9~kpc and is quite metal poor with an average [Fe/H] of $-2.35$~dex \citep[][2010 edition]{Harris_1996}, so it is a fairly extreme test of our pipeline's ability to measure abundances with hot, metal-poor stars on the MSTO\@. A CMD of the stars used in our analysis is shown in Figure~\ref{fig:GC_CMDs} to highlight that this sample includes many stars around the MSTO, which makes this a good test for the pipeline's accuracy for HALO7D-like data. The photometry comes from \citet{Stetson_2019}. The spectral observations have $20 < \mathrm{SNR} < 100~\rm \AA$ and were observed with Keck II/DEIMOS in a similar configuration to the HALO7D data; see \citet{Escala_2019} for details. The main differences between the DEIMOS configurations are that the GC data used a 0.8'' slitwidth and a central wavelength of $\sim 7500~\rm \AA$ instead of $\sim 7200~\rm \AA$\@.

To transform the Johnson-Cousins photometry of \citet{Stetson_2019}, we use the $m_u$ magnitudes and $m_u-m_v$ colors of each cluster and align the literature MIST isochrone on the CMD data using the appropriate filters. With this, we find the closest matching point on the isochrone to each star, which assigns an estimated mass to each star. Using this mass estimate and the same isochrone in HST filters gives each star's approximate $m_{F606W}$ and $m_{F814W}$ magnitudes.

We use the same prior-building process for the GC stars as for the MW halo stars. That is, the known distances, ages, and abundances of the clusters are not used; instead, we assume that all the cluster stars are distributed like MW halo stars. While this leads to larger posterior uncertainties than we could achieve using the known GC parameters, it gives a better measurement of how well the pipeline will perform with MW halo stars. These GC comparisons also function to show the insensitivity of the pipeline to our choice of priors in Table \ref{t:prior_distributions}; we know the prior distributions do not describe the GC stars very well, so our ability to recover reasonable posterior abundances for stars with spectral $\mathrm{SNR} > 20~\rm \AA^{-1}$ is not overly dependent on our priors. 

For each cluster, we measure a population average and scatter in $\feh$ and $\alphafe$ using the posterior distributions of each star. For M92, we measure an average $\feh$ of $-2.31$~dex with a scatter of $0.08$~dex and an average $\alphafe$ of $0.16$~dex with a scatter of $0.10$~dex. For M2, we measure an average $\feh$ of $-1.64$~dex with a scatter of $0.10$~dex and an average $\alphafe$ of $0.19$~dex with a scatter of $0.09$~dex. These means in $\feh$ are consistent with the literature values quoted above. A significant contribution to the scatter is that we force the priors to contain MW halo-like properties instead of the properties of each cluster. This leads to stars with lower SNR spectra relying on incorrect prior information, thereby increasing the scatter. 

\begin{figure}[t]
\begin{center}
\begin{minipage}[c]{\linewidth}
\includegraphics[width=\linewidth]{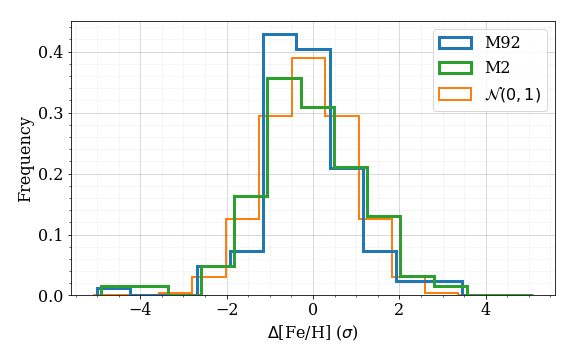}
\end{minipage}\\
\begin{minipage}[c]{\linewidth}
\includegraphics[width=\linewidth]{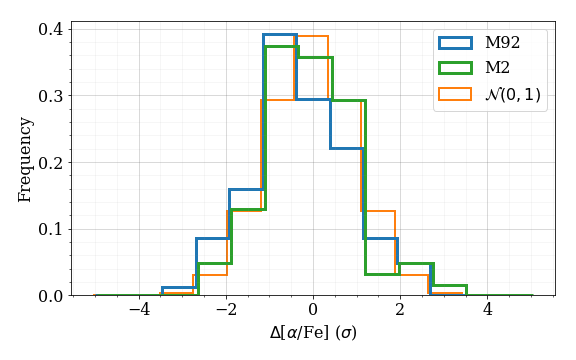}
\end{minipage}
\caption{Uncertainty-scaled comparison of measured $\feh$ and $\alphafe$ with population averages for stars in M92 and M2\@. The green lines show that a unit Gaussian has good agreement with our results.}
\label{fig:GC_chemistry_hists}
\end{center}
\end{figure}

We are most interested in assessing whether the pipeline returns abundance measurements that are unbiased over the HALO7D range of spectral SNR and whether the posterior uncertainties are reasonable. To that end, we compare the chemical abundances to our population averages of $\feh$ and $\alphafe$ for the GCs instead of to literature values. This is shown in Figure~\ref{fig:GC_chemistry_hists}, where the uncertainty-scaled $\feh$ and $\alphafe$ distributions for both clusters agree with the unit Gaussian. This is evidence that the pipeline is giving reasonable abundances for real stars with spectra like those of HALO7D\@.

In addition to not being MW halo stars, and therefore not truly following the assumptions that go into the priors, the cluster data is dissimilar to HALO7D in a few other ways. First, all the GC data consists of a single spectroscopic observation per star instead of multiple, which means that any systematic errors in an individual spectrum are more likely to impact the results (e.g. poor wavelength solution, bad skyline removal, vignetting). Second, the GC photometry comes from ground-based observations which have significantly larger uncertainties than our HALO7D photometry measured with HST; this large color uncertainty leads to significantly more diffuse priors for the GC stars as compared to HALO7D stars. Third, the GC spectra have wavelength solutions that are generally less well-behaved than HALO7D spectra because the observations were calibrated with a single set of relatively red arc lamp exposures and didn't use a second set of bluer arc lamps as was done for the HALO7D observations. Finally, the DEIMOS configuration for the GC stars is centered on $\sim7500~\rm \AA$ instead of the $\sim7200~\rm \AA$ used in HALO7D, which means that many of the GC spectra do not extend to wavelengths as blue as $\sim 5000~\rm \AA$ where many of the strongest $\feh$ and $\alphafe$ features exist. 

\section{Statistical Significance of HALO7D Anisotropy Differences} \label{sec:anisotropy_pipeline}

To test the statistical significance of the differences we see in the anisotropy measurements of the four HALO7D fields, we generate simulated HALO7D-like surveys and measure their kinematics. For each HALO7D-like realization, we assume that each of the four fields has the same number of stars as we observed: 40 in EGS, 36 in COSMOS, 21 in GOODSN, and 16 in GOODSS\@. For the fraction of disk contamination, we again use the values measured in our HALO7D analysis, such that 2 of the EGS stars are disk contaminants, as are 2 of the GOODSN and GOODSS stars, and 1 of the COSMOS stars; the remaining 106 of the 113 stars are halo stars.

\begin{figure}[h]
\begin{center}
\includegraphics[width=\linewidth]{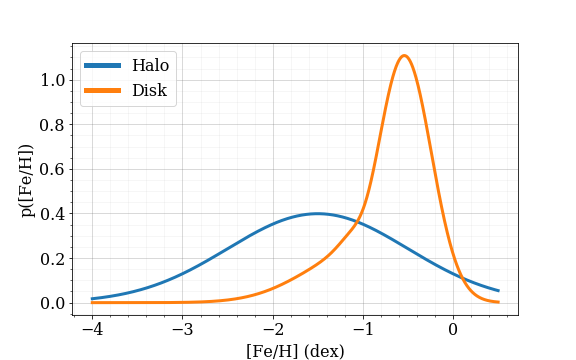}
\caption{Comparison of the chosen prior Halo and Disk model MDFs\@. The Halo model is a simple Gaussian centered at $-1.5$~dex with a scatter of $1.0$~dex while the Disk model is a mixture of a Gaussian and skew-normal distribution as described in Equation \ref{eq:disk_mdf}. The Disk's distribution comes from an analytical approximation of the `high-alpha disk'' and ``metal-weak thick disk'' populations of \citet{Naidu_2020}.}
\label{fig:prior_mdf_comparison}
\end{center}
\end{figure}

We generate realizations of HALO7D that follow the same magnitude and color limits in each of the four fields. For the disk stars, the $\feh$s are drawn from
\begin{equation} \label{eq:disk_mdf}
    \begin{split}
        \feh \sim &\frac{1}{6}\mathcal{SKN}(\mu=-1.05~\mathrm{dex},\\
                  &\hspace{1.75cm}\sigma=0.6~\mathrm{dex},a=-5)\\
                  &+\frac{5}{6}\mathcal{N}(\mu=-0.54~\mathrm{dex},\sigma=0.3~\mathrm{dex})
    \end{split}
\end{equation}
the ages are drawn from
\begin{equation} \label{eq:disk_age_dist}
    \begin{split}
        \age \sim \mathcal{N}(\mu=10~\mathrm{Gyr},\sigma=2~\mathrm{Gyr})
    \end{split}
\end{equation}
and the masses and distance moduli are drawn from the prior distributions shown in Table \ref{t:prior_disk_distributions}. The $\feh$ distribution comes from an analytical approximation of the ``high-alpha disk'' and ``metal-weak thick disk'' populations of \citet{Naidu_2020}, the age distribution comes from another analytical approximation of the ``high-alpha disk'' from \citet{Bonaca_2020}. The resulting prior MDFs for the Halo and Disk models are compared in Figure \ref{fig:prior_mdf_comparison}.

The velocity components for the disk stars are drawn from the prior distributions in the bottom half of Table \ref{t:prior_velocity_distributions}. For the halo stars, the $\feh$, age, mass, and distance modulus are drawn from the prior distributions shown in Table \ref{t:prior_distributions}. For the kinematics of each halo star, we assume that all four fields have the same halo velocity ellipsoid that matches the values we measure for the total HALO7D population. Specifically, $\langle v_r \rangle = \langle v_\theta \rangle = 0~{\rm km \ s}^{-1}$, $\langle v_\phi \rangle = -47~{\rm km \ s}^{-1}$, and $(\sigma_r, \sigma_\phi, \sigma_\theta) = (125,70,80)~{\rm km \ s}^{-1}$, which implies an anisotropy of $\beta = 0.57$\@. The velocity components for each halo star are drawn from the distributions in the top half of Table \ref{t:prior_velocity_distributions}.

With each star having a defined $\feh$, age, mass, and distance modulus, we use the corresponding MIST isochrone to get the apparent $m_{F606W}$ and $m_{F814W}$ magnitudes, ensuring that the colors and magnitudes are within the limits that we observe for each HALO7D field. The distance to each star is used to transform the velocity components into proper motions and LOS velocities, which means that each fake star has a set of observables that make them comparable to the HALO7D sample. Each proper motion component is given an uncertainty equal to the median uncertainty from the HALO7D sample: $\sigma_{\mu_l\cdot \cos b} = 0.17~{\rm mas \ year}^{-1}$, $\sigma_{\mu_b} = 0.16~{\rm mas \ year}^{-1}$\@. For the LOS velocities, we use Figure 7 of \citet{Cunningham_2019a}, which shows the relationship between $\sigma_{V_{LOS}}$ and a star's apparent $m_{606W}$ magnitude for the HALO7D sample; we capture this relationship as 
\begin{equation*}
    \begin{split}
        \sigma_{V_{LOS}}(m_{F606W}) =& (4\times 10^{-11}~{\rm km \ s}^{-1})\cdot\exp\left(\frac{m_{F606W}}{0.86859~\mathrm{mag}}\right)+1.5~{\rm km \ s}^{-1}
    \end{split}
\end{equation*}
which gives LOS velocity uncertainties of $1.6~{\rm km \ s}^{-1}$ for $m_{606W} = 19$~mag,  $5.5~{\rm km \ s}^{-1}$ for $m_{606W} = 22$~mag, and $41.5~{\rm km \ s}^{-1}$ for $m_{606W} = 24.5$~mag. 

\begin{figure}[h]
\begin{center}
\includegraphics[width=\linewidth]{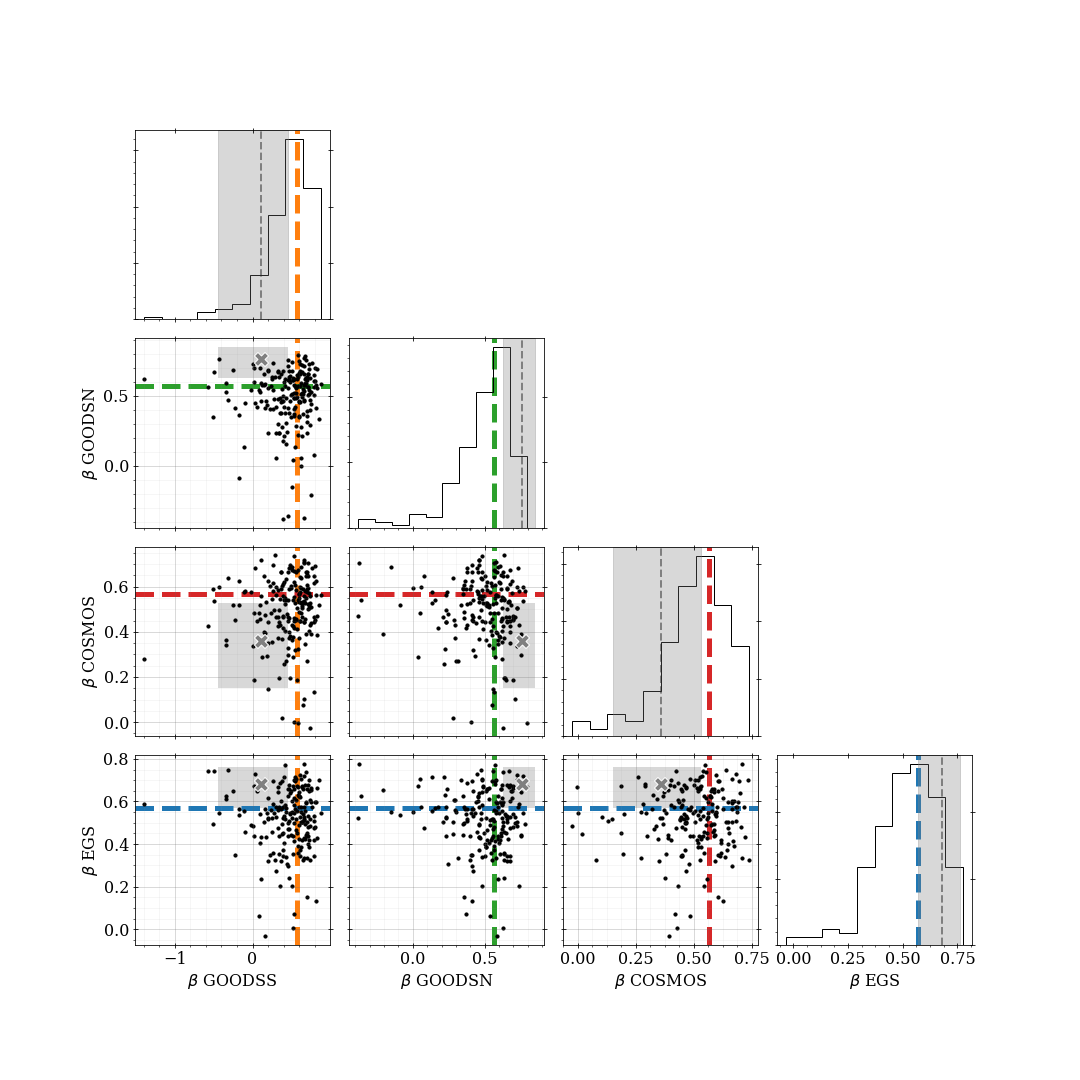}
\caption{Distributions of median anisotropy $\beta$ for 129 HALO7D-like realizations for each field. The black points show the posterior median $\beta$ for each realization, and the histograms on the diagonal show their 1D distribution. The dashed vertical colored lines show the input $\beta$ that was used to generate the data. The grey dashed lines in the histograms and the grey Xs in the scatter plots show the posterior median $\beta$ we measure for each field in the real HALO7D data. The grey-shaded regions show the 68\% region of the $\beta$ distribution for the real data. There are only 3 black points that are within the 4D hypercube defined by the grey-shaded region.  This implies that the differences in $\beta$ that we measure for the real data are not likely explained by random chance. }
\label{fig:anisotropy_testing_results}
\end{center}
\end{figure}

As with the HALO7D sample, we generate absolute magnitude prior distributions assuming each star belongs to the Thick Disk and the Halo. With the absolute magnitude distributions giving corresponding distance distributions, we are able to follow the same steps as outlined in Section \ref{sec:discussion}, fitting for the halo velocity components ($\langle v_\phi \rangle$, $\sigma_r$, $\sigma_\phi$, $\sigma_\theta$) and thick disk fraction, $f_{disk}$\@. We repeat this process for each realization of HALO7D and measure the median $\beta$ for each of the simulated fields. We keep results from the realizations that had a measured $\beta$ for the total sample that is within the $68\%$ region of the real data's total sample (i.e. only cases where the fake star total sample's median $\beta$ agrees closely with the data's $\beta_{\mathrm{TOTAL}}=0.57_{-0.08}^{+0.07}$); for our 200 realizations, we find 129 of the medians fall within this region, which agrees with the expectation from a binomial distribution with $p=0.68$ and $n=200$. These results are summarized in Figure \ref{fig:anisotropy_testing_results}, where the black points show the median $\beta$ measured for each of the 129 realizations. The dashed colored lines show the input anisotropy, $\beta = 0.57$, while the shaded grey regions show the 68\% region of the results we measure for the real HALO7D data. The dashed grey lines and grey Xs show the median $\beta$ we measure for the real HALO7D data. 

From the 129 simulated HALO7D realizations with total $\beta$ in agreement with the input value, we find only 3 cases that are similar to the results we measure from the real HALO7D data. In particular, there are 3 realizations where the measured $\beta$ is within the 68\% shaded regions in all four fields. The fractions of medians within each field's 68\% region separately are approximately 31\%, 26\%, 48\%, and 38\% for GOODSS, GOODSN, COSMOS, and EGS respectively. This works out to a 1.5\% probability of seeing data in all four shaded regions at the same time from chance alone, which is similar to the 3/129 realizations we measure in this region. From this analysis, we reject the null hypothesis that the anisotropy differences we see between the four HALO7D fields are likely to be explained by statistical chance. 

\begin{figure}[h]
\begin{center}
\includegraphics[width=\linewidth]{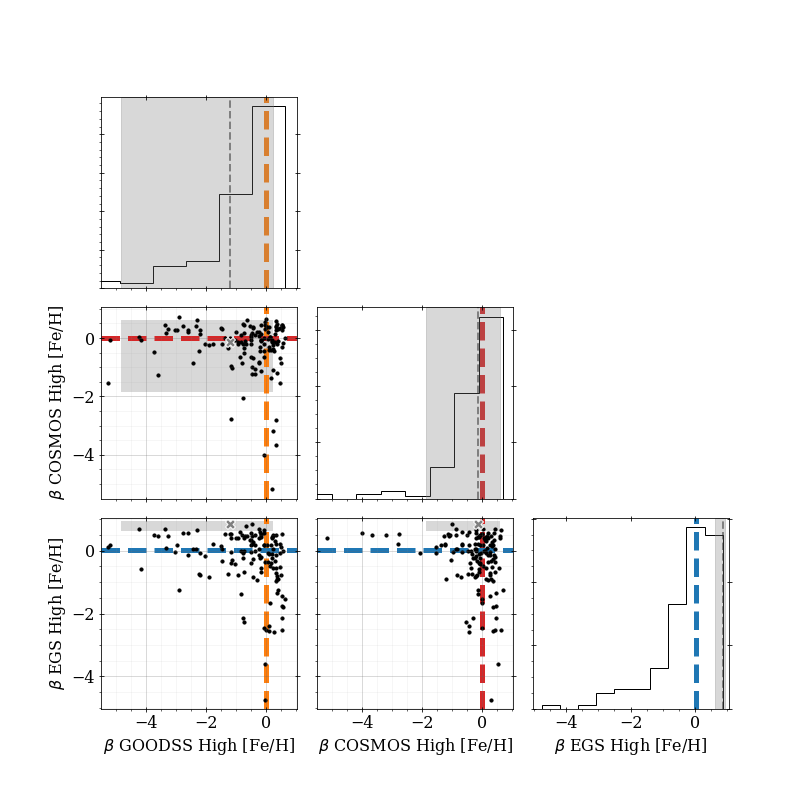}
\caption{Same as Figure \ref{fig:anisotropy_testing_results}, but for 140 realizations in the high $\feh$ bin. The GOODSN panels are omitted because they had too few stars in this $\feh$ bin for a useful analysis. There are only 6 black points that are within the 3D hypercube defined by the grey-shaded region in this figure.  This implies that the differences in $\beta$ that we measure for the real data are not likely explained by random chance.}
\label{fig:anisotropy_testing_results_HIGHFEH}
\end{center}
\end{figure}

We repeat this analysis technique for the different $\feh$ bins. We do this by again assuming that the input velocity components are those measured using the total population in the real HALO7D data in that $\feh$ bin (i.e. the values reported in Table \ref{t:HALO7D_anisotropy_summary}). We again create realizations of HALO7D data, this time drawing the expected number of stars in each of the $\feh$ bins so that the number of stars in each field and each bin matches the totals shown in Table \ref{t:HALO7D_chemistry_summary}. As before, we keep only the $\beta$ medians from the realizations that agree with the 68\% region of the $\beta_{\mathrm{TOTAL}}$ for that $\feh$ bin; this works out to 140, 139, and 130 of the 200 total realizations for the high, mid, and low $\feh$ bins respectively. For the high $\feh$ bin case, the $\beta$ medians of the realizations are shown in Figure \ref{fig:anisotropy_testing_results_HIGHFEH}. We find that $\beta$ medians fall in all the grey shaded regions only 3.3\%, 18\%, and 35\% of the time for the high, mid, and low $\feh$ bins respectively; as with the real HALO7D data, we omit the GOODSS low $\feh$ bin measurements as well as the GOODSN high and low $\feh$ bin measurements for having too few stars. From these probabilities, we can see that the anisotropy distributions we measure in the high $\feh$ bin aren't likely to be produced by random chance alone, meaning the GOODSS, COSMOS, and EGS fields likely have different average halo properties in this $\feh$ bin. For the mid and low $\feh$ bins, these differences are not as statistically significant. In the case of the mid $\feh$ bin, this is in line with our expectations because the fields have similar $\beta$ measurements in the HALO7D data, suggesting that the velocity distributions of the stars in this $\feh$ bin are more similar between the fields. For the low $\feh$ bin, these results tell us that the differences in anisotropy we see between COSMOS and EGS in this bin could be produced by chance alone approximately $\sim2/5$ of the time. 

\end{document}